\documentclass[11pt]{article}

\usepackage{epsfig,subfigure}
\usepackage{amssymb, amsmath}
\usepackage{color}
\usepackage{graphicx}
\usepackage{color}

\usepackage{amssymb}
\usepackage{amsmath,amsfonts,amssymb}
\usepackage{verbatim}
\usepackage{stfloats}
\usepackage[bookmarks=false]{}

\newtheorem{theorem}{Theorem}

\newcommand\blfootnote[1]{%
  \begingroup
  \renewcommand\thefootnote{}\footnote{#1}%
  \addtocounter{footnote}{-1}%
  \endgroup
}

\begin{document}
\date{}
\title{Rank Matching for Multihop Multiflow}
\author{ \normalsize Hua Sun, Sundar R. Krishnamurthy and Syed A. Jafar \\
}
\maketitle

\blfootnote{This work will be presented in part at GLOBECOM 2014. Hua Sun (email: huas2@uci.edu), Sundar R. Krishnamurthy (email: srkrishn@uci.edu) and Syed A. Jafar (email: syed@uci.edu) are with the Center of Pervasive Communications and Computing (CPCC) in the Department of Electrical Engineering and Computer Science (EECS) at the University of California Irvine. }

\begin{abstract}
We study the degrees of freedom (DoF) of the layered $2\times 2\times 2$ MIMO interference channel  where each node is equipped with  arbitrary number of antennas,  the channels between the nodes have arbitrary rank constraints, and subject to the rank-constraints the channel coefficients can take arbitrary values. The DoF outer bounds reveal a fundamental rank-matching phenomenon, reminiscent of impedance matching in circuit theory.  It is well known that the maximum power transfer in a circuit is achieved not for the  maximum or minimum load impedance but for the load impedance that matches the source impedance. Similarly, the maximum DoF in the rank-constrained $2\times 2 \times 2$ MIMO interference network is achieved not for the maximum or minimum ranks of the destination hop, but when the ranks of the destination hop match the ranks of the source hop. In fact, for mismatched settings of interest, the outer bounds identify a DoF loss penalty that is  precisely equal to the rank-mismatch between the two hops. For symmetric settings, we also provide achievability results to show that along with the min-cut max-flow bounds, the rank-mismatch bounds are the best possible, i.e., they hold for \emph{all} channels that satisfy the rank-constraints and are tight for \emph{almost all} channels that satisfy the  rank-constraints. Limited extensions --- from sum-DoF to DoF region, from 2 unicasts to $X$ message sets, from 2 hops to more than 2 hops and from 2 nodes per layer to more than 2 nodes per layer --- are considered to illustrate how the insights generalize beyond the elemental $2\times 2\times 2$ channel model.

\end{abstract}
\newpage
\allowdisplaybreaks
\section{Introduction}\label{section:introduction} 
The $2\times 2\times 2$ interference channel, which is a layered network comprised of two source nodes, two relay nodes and two destination nodes, is an elemental model for the study of the information theoretic foundations of multihop multiflow networks. Many of the key ideas behind multihop multiflow networks, such as interference neutralization \cite{Mohajer_Diggavi_Fragouli_Tse}, aligned interference neutralization \cite{Gou_Wang_Jafar_Jeon_Chung}, aligned interference diagonalization \cite{Shomorony_Avestimehr_cake}, opportunistic scheduling \cite{Jeon_Chung_Jafar}, network condensation and manageable interference \cite{Shomorony_Avestimehr, Wang_Gou_Jafar_MU}  have been discovered through the degrees of freedom (DoF) studies of the $2\times 2\times 2$ interference channel and its natural extensions to more than 2 sources/relays/destinations/hops, arbitrary topologies, and even non-layered settings \cite{Gou_Wang_Jafar_Nonlayer}. Continuing along this path, in this work we explore a generalization of the $2\times 2\times 2$ interference network to the multiple-input-multiple-output (MIMO) setting with arbitrary ranks for each of the channels involved. The goal  is to shed light on the information theoretic implications of the dimensionality constraints of the sub-networks comprising a multihop multiflow network. Parameterizing the problem in terms of the ranks of each of the constituent channels, allows us to go beyond the basic min-cut arguments  to identify an intriguing ``\emph{rank matching}" property, somewhat reminiscent of ``\emph{impedance matching}" in circuit theory. It is well known that the maximum power transfer in a circuit is achieved not for the  maximum or minimum load impedance but for the load impedance that matches the source impedance. Similarly, the maximum DoF in the elementary $2\times 2 \times 2$ MIMO interference network is achieved not for the maximum or minimum ranks of the destination hop, but when the ranks of the destination hop match the ranks of the source hop. In fact, for mismatched settings of interest, the loss in DoF turns out to be  precisely equal to the rank-mismatch between the two hops.

\begin{figure}[h]
\begin{center}
\includegraphics[scale=0.55]{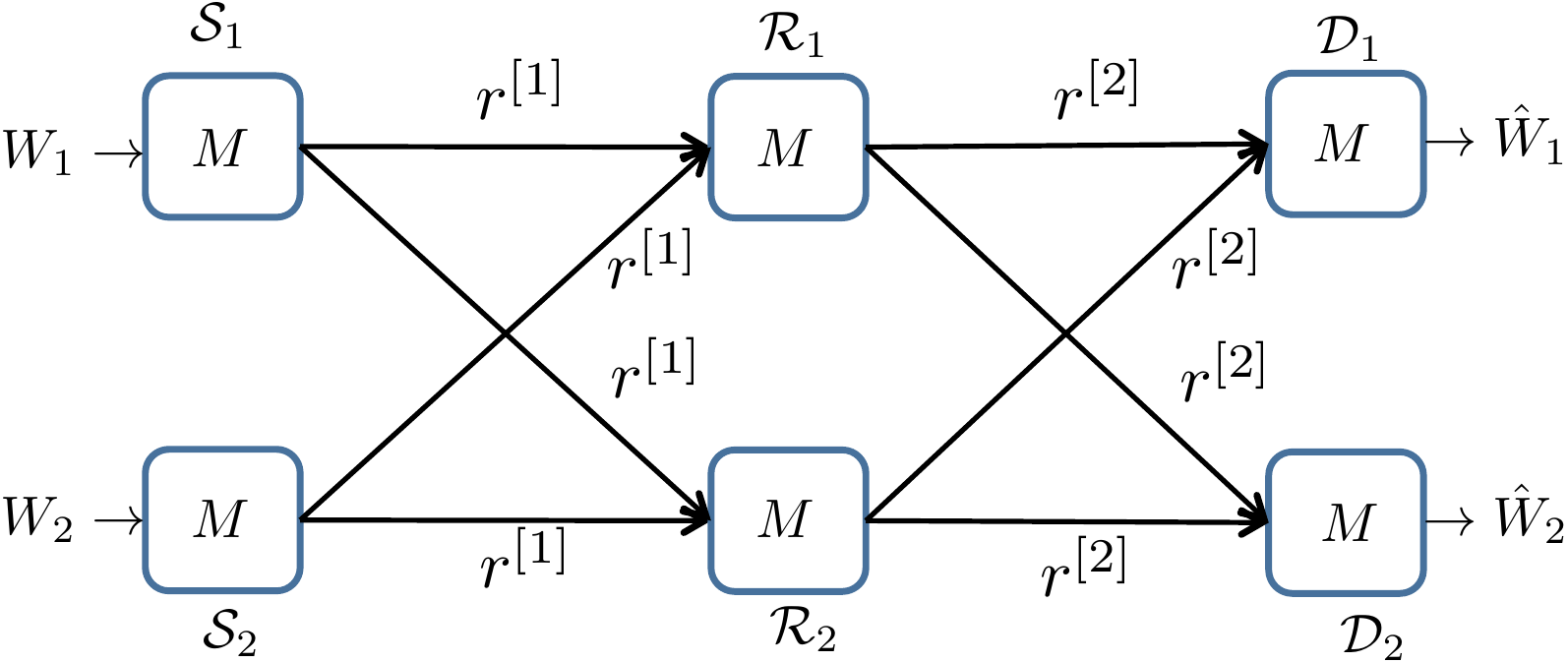}
\caption{\small $2\times 2\times 2$ MIMO interference channel with $M$ antennas at each node where all channels in the first hop have rank $r^{[1]}$ and all channels in the second hop have rank $r^{[2]}$.}\label{fig:2x2x2}
\end{center}
\end{figure}
As an example, consider the $2\times 2\times 2$ MIMO interference channel illustrated in Fig. \ref{fig:2x2x2} where all nodes are equipped with $M$ antennas, all channels in the first hop have rank $r^{[1]}$, and all channels in the second hop have rank $r^{[2]}$. Aside from the rank-constraints, the channels can take arbitrary  values. The min-cut max-flow bound for this network simply states that the sum-DoF, $d_\Sigma\leq\min(4r^{[1]},4r^{[2]}, 2M)$. However, as we show in this work, the rank-constraints enforce the following rank-mismatch bound on the sum-DoF.
\begin{eqnarray}
d_\Sigma&\leq&2M-\Delta r
\end{eqnarray}
where $\Delta r = |r^{[1]}-r^{[2]}|$ is the rank-mismatch term. Combined with the min-cut max-flow bounds, this produces the tightest possible bound for the given rank-constraints,
\begin{eqnarray}
d_\Sigma&\leq&\min (4r^{[1]},4r^{[2]}, 2M-\Delta r)
\end{eqnarray}
This is the tightest bound possible in the sense that  1) it holds for \emph{all} channels that satisfy the  given rank-constraints, and 2) there exist channels that satisfy the given rank-constraints for which the bound is tight. In fact, the bound is tight for \emph{almost all} channels that satisfy the rank-constraints. Remarkably, except for severely rank-deficient scenarios when the min-cut max-flow bounds are active, for  moderately rank-deficient settings that are of main interest, it is  the rank-mismatch bound that is active. Also note that the best possible outcome, $d_\Sigma=2M$, sometimes referred to as ``everyone gets the entire cake"  \cite{Gou_Wang_Jafar_Jeon_Chung, Jeon_Chung_Jafar, Shomorony_Avestimehr_cake}, is possible only if $\Delta r=0$, i.e., ranks in the two hops are matched.

\begin{figure}[h]
\begin{center}
\includegraphics[scale=0.55]{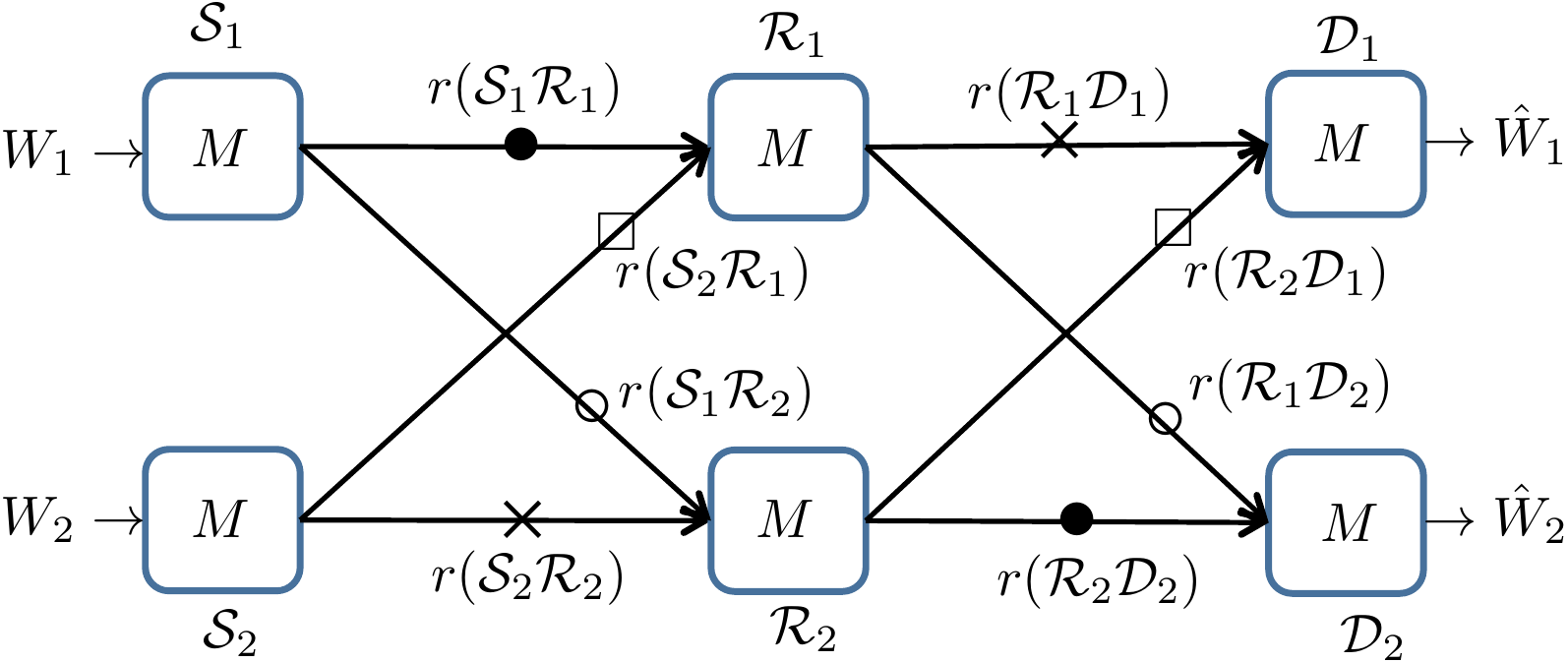}
\caption{\small A $2\times 2\times 2$ MIMO interference channel with $M$ antennas at each node and arbitrary ranks. The symbols on the links identify the pairing of channels  in rank-matching outer bounds. $2M$ DoF are not achievable unless similarly marked channels  have the same ranks, e.g., $r(\mathcal{S}_1\mathcal{R}_1)$ must be equal to $r(\mathcal{R}_2\mathcal{D}_2)$.}\label{fig:2x2x2general}
\end{center}
\end{figure}
The rank matching phenomenon is not limited to symmetric settings. Consider, for example the case illustrated in Fig. \ref{fig:2x2x2general} where all ranks are allowed to be different from each other. The rank-mismatch bound here takes the following form.
\begin{align}
d_\Sigma&\leq 2M-\Delta r\\
\Delta r&= \max\left(|r(\mathcal{S}_1\mathcal{R}_1)-r(\mathcal{R}_2\mathcal{D}_2)|,|r(\mathcal{S}_2\mathcal{R}_2)-r(\mathcal{R}_1\mathcal{D}_1)|, |r(\mathcal{S}_1\mathcal{R}_2)-r(\mathcal{R}_1\mathcal{D}_2)|, |r(\mathcal{S}_2\mathcal{R}_1)-r(\mathcal{R}_2\mathcal{D}_1)|\right)\nonumber
\end{align}
Note that in terms of rank-matching, a channel originating at $\mathcal{S}_1(\mathcal{S}_2)$ is paired with a channel terminating at $\mathcal{D}_2(\mathcal{D}_1)$, and a channel terminating at $\mathcal{R}_1(\mathcal{R}_2)$ is paired with a channel originating at $\mathcal{R}_2(\mathcal{R}_1)$. The channel pairings are indicated in Fig. \ref{fig:2x2x2general}. $2M$ DoF cannot be achieved unless each of these pairs of associated channels have matching ranks. 

\begin{figure}[h]
\begin{center}
\includegraphics[width=3in]{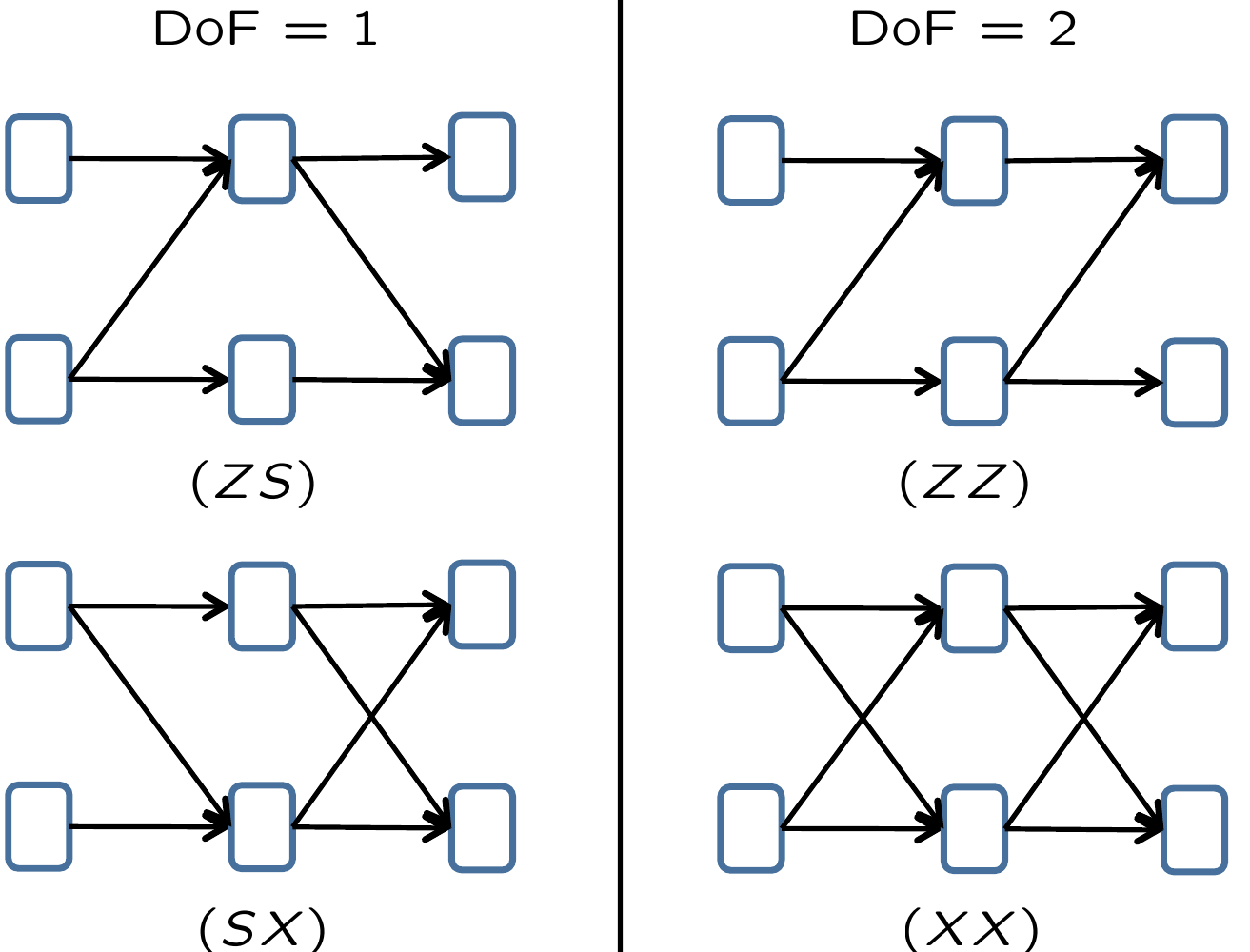}
\caption{\small Some topologies of $2 \times 2 \times 2$ SISO interference channel. $ZS$ and $SX$ topologies have DoF 1 as $r(\mathcal{S}_2\mathcal{R}_1)$ does not match $r(\mathcal{R}_2\mathcal{D}_1)$. In contrast, all ranks match in $ZZ$ and $XX$ topologies such that they have 2 DoF.}
\label{siso}
\end{center}
\end{figure}

As a simple application of this bound, let us recover the DoF results for the various non-trivial topologies of the $2\times 2\times 2$ SISO ($M=1$) interference channel. Following the terminology of  \cite{Mohajer_Diggavi_Fragouli_Tse}, these are  labeled as  the $ZZ$, $SS$, $ZS$, $SZ$, $XZ$, $XS$, $ZX$, $SX$ and $XX$ topologies. Fig. \ref{siso} illustrates some of them. The rank-mismatch bound immediately identifies  $SZ, ZS, SX, ZX, XS, XZ$ as the rank mis-matched topologies $(\Delta r=1)$ which can therefore only have 1 DoF, whereas $SS, ZZ, XX$ are the rank-matched topologies $(\Delta r=0)$, which have indeed been shown to have $2$ DoF.

The rank matching phenomenon persists even in further generalized settings  with arbitrary antenna configurations and/or redundant dimensions, i.e., when certain signal dimensions at a  node may be inaccessible to/from any other node. Indeed, to gain as much insight as possible, we consider the generalized setting in this work. This is described in the system model that we present next.

\section{System Model}\label{sec:systemmodel} 
The $2 \times 2 \times 2$ MIMO interference channel is comprised of 3 layers and there are two nodes in each layer. Layer 1 contains the two source nodes $\mathcal{S}_1, \mathcal{S}_2$, layer  2 contains the two relay nodes $\mathcal{R}_1, \mathcal{R}_2$, and layer 3 contains the two destination nodes, $\mathcal{D}_1, \mathcal{D}_2$. The $j$-th source, relay, and destination node is equipped with $M(\mathcal{S}_j), M(\mathcal{R}_j), M(\mathcal{D}_j)$ antennas, respectively. In addition to this notation which identifies the sources, relays and destinations explicitly and is therefore easier to grasp, we will also use an alternative compact notation which identifies nodes only by the layer index when brevity is the priority, e.g., in the details of the longer proofs. According to this compact notation, the $j$-th node in layer $l$ has $M^l_j$ antennas, $j \in \{1,2\}, l \in \{1,2,3\}$. So, for example, $M(\mathcal{R}_2)=M_{2}^2$ and $M(\mathcal{D}_1)=M_1^3$. 

At time index $t\in\mathbb{N}$, the various inputs and outputs are related as follows.
\begin{eqnarray}
{\bf Y}_{j}^{l+1}(t) &=& \sum_{i=1}^2 {\bf H}_{ji}^{l} (t) {\bf X}_{i}^{l} (t)  + {\bf Z}_j^{l+1}(t), ~~~ j \in \{1,2\}, l \in \{1,2\}
\end{eqnarray}
where ${\bf Y}_{j}^{l+1} (t)$ is the $M_j^{l+1} \times 1$ received signal vector observed at node $j$ in layer $l+1$, ${\bf X}_{i}^{l} (t)$ is the $M_i^{l} \times 1$ transmitted signal vector sent by node $i$ in layer $l$ and ${\bf Z}_{j}^{l+1} (t)$ is the $M_j^{l+1} \times 1$ vector of independent and identically distributed (i.i.d.)  zero mean unit variance circularly symmetric complex Gaussian noise terms, respectively. ${\bf H}_{ji}^{l}(t)$ is the $M_j^{l+1} \times M_i^l$ channel matrix from node $i$ in layer $l$ to node $j$ in layer $l+1$. In other words, ${\bf H}_{ji}^{l}(t)$ is the channel matrix between node $i$ and node $j$ over the $l$-th hop. All symbols are complex and noise processes are i.i.d over time. $\mathcal{S}_i$ has an independent message $W_i$ for $\mathcal{D}_i$, $i \in \{1,2\}$. Each transmitting node is subject to average power constraint $P$.  The encoding functions at the relays are assumed to be known everywhere.
The time index, $t$, will occasionally be suppressed for concise notation, when no ambiguity would be caused. 

The rank-constraints are stated as follows, $\forall t\in \mathbb{N}. $
\begin{eqnarray}
\begin{array}{llll}
\text{rank}({\bf H}_{11}^1(t)) =  r(\mathcal{S}_1\mathcal{R}_1)&\text{rank}({\bf H}_{12}^1(t)) =  r(\mathcal{S}_2\mathcal{R}_1)&\text{rank}({\bf H}_{21}^1(t)) =  r(\mathcal{S}_1\mathcal{R}_2)&\text{rank}({\bf H}_{22}^1(t)) =  r(\mathcal{S}_2\mathcal{R}_2)\\
\text{rank}({\bf H}_{11}^2(t)) =  r(\mathcal{R}_1\mathcal{D}_1)&\text{rank}({\bf H}_{12}^2(t)) =  r(\mathcal{R}_2\mathcal{D}_1)&\text{rank}({\bf H}_{21}^2(t)) =  r(\mathcal{R}_1\mathcal{D}_2)&\text{rank}({\bf H}_{22}^2(t)) =  r(\mathcal{R}_2\mathcal{D}_2)
\label{eq:rank}
\end{array}
\end{eqnarray}

The channel coefficients can take arbitrary values  and are also allowed to vary in time as long as the rank-constraints are satisfied and the non-zero singular values of each channel matrix are bounded away from zero and infinity. Unless stated explicitly, we do not require that the channels be in general position. Perfect channel knowledge is assumed everywhere.  Finally, the definitions of codebooks, achievable rates, capacity, and degrees of freedom are all used  in the standard sense.

\section{Results}
In this section we present our two main results --- the general statement of the rank mismatch outer bound, and a proof that (along with the min-cut max-flow bound) it is tight, at least in symmetric settings.
\subsection{Rank-Mismatch Outer Bound}
Without loss of generality, let us discard any redundant dimensions (dimensions that are not accessible to/from any other node) from the sources and destinations, respectively, so that,
\begin{eqnarray}
M({\mathcal{S}_i}) &\leq& r(\mathcal{S}_i\mathcal{R}_1) + r(\mathcal{S}_i\mathcal{R}_{2}),~~ i \in \{1,2\}\\
M({\mathcal{D}_k}) &\leq&r(\mathcal{R}_1\mathcal{D}_k) + r(\mathcal{R}_2\mathcal{D}_{k}),~~ k \in \{1,2\}
\end{eqnarray}
Similarly discarding redundant dimensions at the relays,  the effective number of transmit antennas $M_t({\mathcal{R}_j})$, and the effective number of receive antennas $M_r({\mathcal{R}_j})$ at the $j$-th relay, $j\in\{1,2\}$, are constrained as follows.
\begin{eqnarray}
M_t({\mathcal{R}_j}) &\leq & r(\mathcal{R}_j\mathcal{D}_1) + r(\mathcal{R}_j\mathcal{D}_{2})\\
M_r({\mathcal{R}_j}) &\leq& r(\mathcal{S}_1\mathcal{R}_j) + r(\mathcal{S}_2\mathcal{R}_{j})
\end{eqnarray}

\noindent For compact notation, let us define
\begin{eqnarray}
\bar{i}=\left\{\begin{array}{ll}
1,&\mbox{ if } i=2\\
2,&\mbox{ if } i=1\\
\end{array}\right., &&
\bar{j}=\left\{\begin{array}{ll}
1,&\mbox{ if } j=2\\
2,&\mbox{ if } j=1\\
\end{array}\right.
\end{eqnarray}
With these simplifications of  the notation, we are ready to state the main result in the following theorem.

\begin{theorem}\label{out}
For the rank-constrained $2 \times 2 \times 2$ MIMO interference channel defined in Section \ref{sec:systemmodel}, the sum-DoF, $d_\Sigma$, satisfy the following outer bound for all $i,j\in\{1,2\}$.
\begin{eqnarray}
d_\Sigma&\leq&\frac{1}{2}\left\{\left[M(\mathcal{S}_i)+M_r(\mathcal{R}_j)\right]+\left[M_t(\mathcal{R}_{\bar{j}})+M({\mathcal{D}_{\bar{i}}})\right]\right\}-|\Delta r_{ij}|\label{eq:main}
\end{eqnarray}
where
\begin{eqnarray}
\Delta r_{ij}&=&\left[r(\mathcal{S}_i\mathcal{R}_j)-r(\mathcal{R}_{\bar{j}}\mathcal{D}_{\bar{i}})\right]- \frac{1}{2}\left\{\left[M(\mathcal{S}_i)+M_r(\mathcal{R}_j)\right]-\left[M_t(\mathcal{R}_{\bar{j}})+M({\mathcal{D}_{\bar{i}}})\right]\right\}
\end{eqnarray}
%
\end{theorem}

{\it Remark:} Note that the bounds have a dual character, i.e., the same bounds hold for the reciprocal network obtained by reversing the direction of communication.

{\it Remark:} Note that for all $i,j\in\{1,2\}$,  the first hop channel $\mathcal{S}_i\mathcal{R}_j$ is paired with the second hop channel $\mathcal{R}_{\bar{j}}\mathcal{D}_{\bar{i}}$. This is the same pairing as indicated in Fig. \ref{fig:2x2x2general}. In the best case scenario, the rank-mismatch bound that is active  is the average of the  number of antennas in the two paired channels. This best case corresponds to  the rank-mismatch term $\Delta r_{ij}$ taking zero value, which happens only if the difference of ranks between the paired channels equals half of the corresponding difference of the number of antennas. 
\begin{eqnarray}
r(\mathcal{S}_i\mathcal{R}_j)-r(\mathcal{R}_{\bar{j}}\mathcal{D}_{\bar{i}})&=&\frac{1}{2}\left\{\left[M(\mathcal{S}_i)+M_r(\mathcal{R}_j)\right]-\left[M_t(\mathcal{R}_{\bar{j}})+M({\mathcal{D}_{\bar{i}}})\right]\right\}
\end{eqnarray}
The insight obtained here is that ideally the difference of ranks should be half of the difference of antennas in the paired channels. Otherwise, the deviation from the ideal value is the loss term associated with each bound. 

Theorem \ref{out} has  profound implications in terms of the rank-matching phenomenon --- in addition to the examples presented in the introduction section, please refer to the extensions in Section \ref{sec:extension} for interesting insights. However, we note that the theorem is obtained based only on arguments that are fairly standard for DoF bounds, similar to, e.g., \cite{Mohajer_Diggavi_Fragouli_Tse}. As such, this is a remarkable case of simple arguments leading to surprising insights. The proof of Theorem \ref{out} is presented in Section \ref{p1}.

\subsection{Tightness of Rank-Mismatch Outer Bounds}
Having presented the rank-mismatch outer bounds in Theorem \ref{out}, we next consider the natural question `How tight are these bounds?'.  This seems to be a  difficult question to answer in full generality due to the abundance of parameters. Nevertheless, for the symmetric setting illustrated in Fig. \ref{fig:2x2x2}, where all channels in the first hop have rank $r^{[1]}$ and all channels in the second hop have rank $r^{[2]}$, and all nodes have $M$ antennas, we are able to prove that (combined with min-cut max-flow bounds) the rank-mismatch bounds are the best possible bounds for the given rank-constraints. By best possible we mean that 1) the bounds are satisfied by all channels that satisfy the rank-constraints, 
and 2) there exist channels that satisfy the given rank-constraints for which the bounds are tight. Not only that, but the bounds are tight for \emph{almost all} channels that satisfy the rank-constraints, i.e., they are tight almost surely for generic channels, where by generic channels we mean that the channels are drawn according to a continuous distribution over the algebraic variety defined by the rank-constraints. For instance, one may assume that each $M \times M$ channel over the $l$-th hop is a product of an $M \times r^{[l]}$ channel matrix and a $r^{[l]}  \times M$ channel matrix, each of which is generated randomly and independently of the others across space and time, according to a continuous distribution. We state this result as the following theorem.
\begin{theorem}\label{in}
For the rank-constrained symmetric $2 \times 2 \times 2$ MIMO interference channel illustrated in Fig. \ref{fig:2x2x2} the sum-DoF outer bound $d_\Sigma\leq\min(4r^{[1]}, 4r^{[2]}, 2M - |r^{[1]}-r^{[2]}|)$ is the best possible for the given rank-constraints. 
For generic time-varying channels, the bound is tight almost surely.
\end{theorem}
The proof is presented in Section \ref{p2}.

Note that the rank-mismatch bounds may no longer be tight if additional structure is imposed, e.g., through additional rank-constraints. However, subject only to the rank-constraints stated in (\ref{eq:rank}), these bounds appear to be the best possible.  In fact, for all the cases that we have considered so far,  we have found these bounds to be the best possible when combined with min-cut max-flow bounds. 

\section{Extensions}\label{sec:extension}
In this section, to catch a glimpse of the implications of the rank-matching bounds beyond the sum-DoF of the $2\times 2\times 2$ MIMO interference channel, we consider a few limited extensions --- beyond sum-DoF to DoF regions, beyond 2 unicasts to general message sets ($X$ setting),  beyond 2 hops to the $2\times 2\times 2 \times 2$  setting and beyond $2$ nodes per layer to the $K\times K\times K$ setting. In particular, we find that the DoF loss due to rank-mismatch may be circumvented, at least in symmetric settings, through expanded message sets and/or expanded number of hops.

\subsection{Beyond Sum-DoF: DoF Region}
The insights from the sum-DoF characterization are sufficient to establish the DoF region for the symmetric setting, which is given by the rank-mismatch sum-DoF bound combined with single user min-cut max-flow bounds. We state this result as the following theorem.

\begin{theorem}\label{region}
For the rank-constrained symmetric $2 \times 2 \times 2$ MIMO interference channel illustrated in Fig. \ref{fig:2x2x2}, with generic time-varying channels, the DoF region is the set of all tuples $(d_1,d_2)$ satisfying
\begin{align}
d_1 + d_2 \leq 2M - |r^{[1]}-r^{[2]}|
\\
d_1 \leq \min(2r^{[1]}, 2r^{[2]}, M) \\
d_2 \leq \min(2r^{[1]}, 2r^{[2]}, M)
\end{align}
\end{theorem}
The proof is presented in Section \ref{p-r}.

\subsection{Beyond 2 unicasts: $X$ Message Setting}
Next we consider the $X$ message setting, where there is an independent message from each source to each destination. We want to characterize the sum-DoF for the symmetric setting. It turns out that the $4$ messages in the network provide enough flexibility to fully exploit the signal space resources such that the rank-mismatch penalty term disappears and the min-cut max-flow bound is achievable.
We state this result as the following theorem.

\begin{theorem}\label{x}
For the rank-constrained symmetric $2 \times 2 \times 2$ MIMO $X$ channel, whose underlying channels are the same as that of Fig. \ref{fig:2x2x2}, but with 4 independent messages, one from each source to each destination, the min-cut max-flow bounds $d_\Sigma\leq\min(4r^{[1]}, 4r^{[2]},2M)$ are achievable for generic time-varying channels almost surely.
\end{theorem}
The proof is presented in Section \ref{p-x}.

\subsection{Beyond 2 hops: $2\times 2\times 2\times 2$ MIMO Interference Channel}
\begin{figure}[h]
\begin{center}
\includegraphics[scale=0.7]{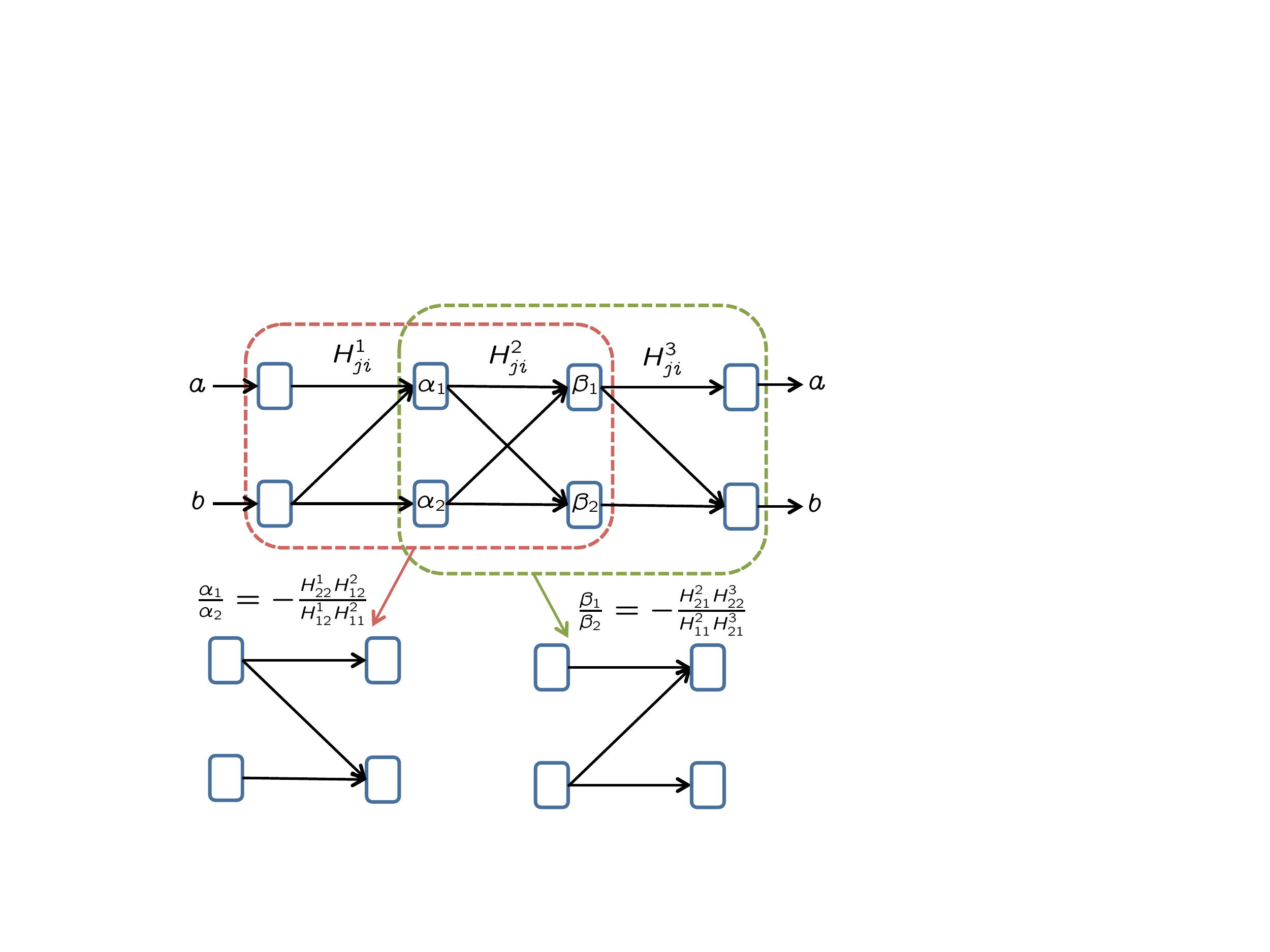}
\caption{\small While the first hop $(Z)$ and the last hop $(S)$ are mismatched if connected directly to each other, linear processing at the intermediate hop facilitates rank-matching so that the first hop, which has a $Z$ topology, sees the rest of the network as a $Z$ topology, and at the same time, the last hop, which has an $S$ topology, sees the rest of the network as an $S$ topology. Thus the presence of the intermediate hop increases the DoF from 1 to 2.
}\label{zxs}
\end{center}
\end{figure}

Consider a 2 unicast  interference network with multiple hops. From the perspective of the relay nodes in any given layer, if linear precoding schemes are employed at all other layers of relay nodes, then the network appears effectively as a $2\times 2\times 2$ MIMO interference network. The rank matching criterion tells us that
from the perspective of the chosen layer of relay nodes, the ranks of the effective channels from the sources to these relays should match the ranks of corresponding channels from these relays to the destinations. Otherwise, Theorem \ref{out} identifies the loss incurred by rank-mismatch. In other words, the goal of other relay layers is to facilitate the matching of ranks as much as possible. This is a useful general design principle and moreover, it is local in the sense that only the net rank information of other hops is needed such that iterative design may be possible. For example, consider the $2 \times 2 \times 2$ SISO interference channel with $ZS$ topology, which has DoF 1. Suppose we are allowed to add a fully connected intermediate hop inside (see Fig. \ref{zxs}), how should we design the relay operations such that we can increase DoF? In this case, it turns out that we can achieve 2 DoF. To see this, let us set
\begin{align}
H_{12}^1H_{11}^2\alpha_1 + H_{22}^1H_{12}^2 \alpha_2= 0 \\
H_{11}^2H_{21}^3\beta_1 + H_{21}^2H_{22}^3 \beta_2= 0 
\end{align}
where $H_{ji}^l$ is the channel coefficient from node $i$ to node $j$ over the $l$-th hop and $\alpha_i, \beta_i$ are the amplify and forward coefficients used by the relays (see Fig. \ref{zxs}). This creates two interference free paths from the sources to their desired destinations. From a rank matching perspective, $\alpha_1,\alpha_2$ are chosen such that the first two hops appear like an $S$ topology to match the last hop, which itself has an $S$ topology, and $\beta_1,\beta_2$ are chosen such that the last two hops appear like a $Z$ topology to match the first hop, which itself has a $Z$ topology. This is illustrated in Fig. \ref{zxs}.

\begin{figure}[h]
\center
\includegraphics[scale=0.55]{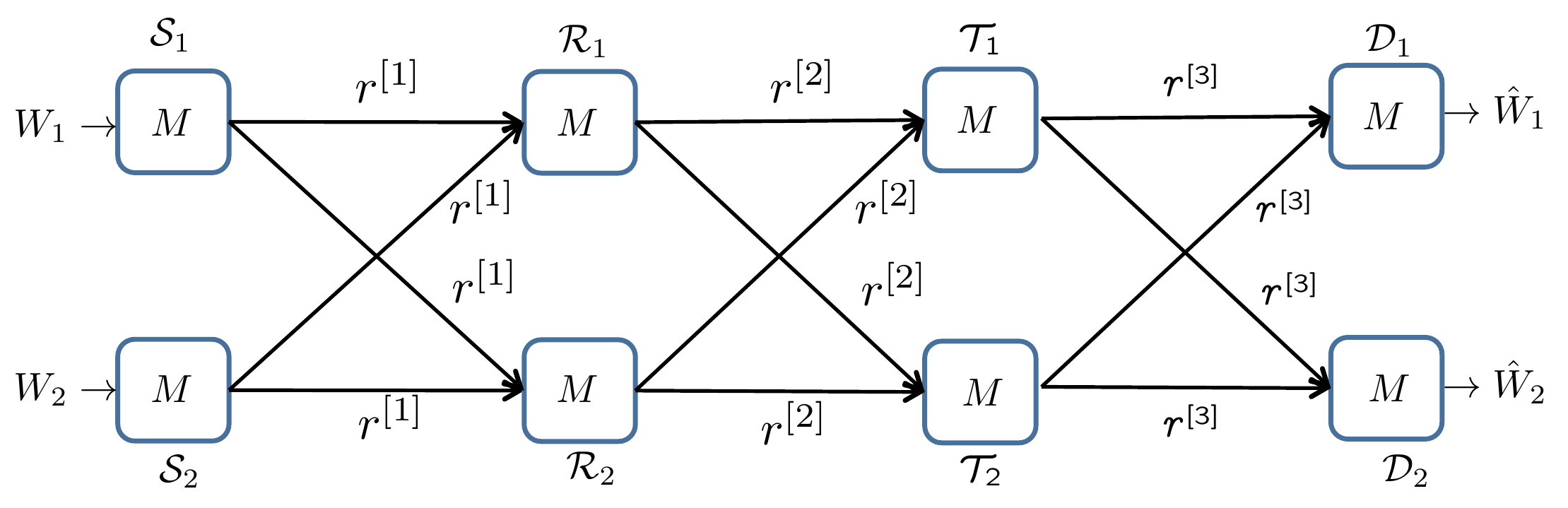}
\caption{\small $2 \times 2 \times 2 \times 2$ MIMO interference channel with $M$ antennas at each node where all channels in the $i$-th hop have rank $r^{[l]}, l \in \{1,2,3\}$.}
\label{3hop}
\end{figure}

{
Motivated by the observation that the intermediate hop can increase DoF by facilitating rank-matching, we explore how much gain can be obtained in the symmetric $2 \times 2 \times 2 \times 2$ MIMO interference channel illustrated in Fig. \ref{3hop}, where all nodes are equipped with $M$ antennas, and all channels in the $l$-th hop have rank $r^{[l]}, l \in \{1,2,3\}$.
Somewhat surprisingly, we show that the min-cut max-flow bounds are tight, for almost all channels that satisfy the rank-constraints. In other words, no matter how much mismatched are the first hop and last hop, the intermediate hop is able to compensate this rank-mismatch, up to its capability, i.e., its own min-cut. That is, 
when the first hop is directly connected to the third hop, the sum-DoF value is $\min(4r^{[1]}, 4r^{[3]}, 2M - |r^{[1]}-r^{[3]}|)$ and if we add the intermediate hop, the sum-DoF value becomes $\min(4r^{[1]}, 4r^{[2]}, 4r^{[3]}, 2M)$ such that the rank-mismatch penalty term disappears. As a result, when $4r^{[2]} \geq \min(4r^{[1]}, 4r^{[3]}, 2M - |r^{[1]}-r^{[3]}|)$, this translates to a strict DoF increase.
We state this result as the following theorem.

\begin{theorem}\label{3hopin}
For the rank-constrained symmetric $2 \times 2 \times 2 \times 2$ MIMO interference channel illustrated in Fig. \ref{3hop} the min-cut max-flow bounds $\min(4r^{[1]}, 4r^{[2]}, 4r^{[3]}, 2M)$ are achievable for generic time-varying channels almost surely.
\end{theorem}
The proof is presented in Section \ref{p3}.

%
}

\subsection{Beyond 2 nodes per layer: $K\times K\times K$ interference network}
Next we consider a case with more than 2 flows, that is, the $K$-DoF feasibility condition for the $K \times K \times K$ interference network, obtained very recently in Theorem 2 of \cite{Ilan_Salman}\footnote{Reference \cite{Ilan_Salman}, which appeared on ArXiv (April 19, 2014) a few weeks after our Globecom submission of this work (March 31, 2014),   independently obtains outer bounds that are similar to our outer bounds,  underscoring the fundamental significance of these bounds.}. While a $K \times K \times K$  network appears to be an  extension that goes beyond the $2 \times 2 \times 2$ interference network that we study here,  we will show that the outer bound needed for the $K$-DoF feasibility result of \cite{Ilan_Salman} also  follows directly from Theorem \ref{out} (sufficiency is also proved for generic channel coefficients in \cite{Ilan_Salman}). This is because clustering nodes (allowing cooperation among them) reduces a $K\times K\times K$ SISO interference network to a $2\times 2\times 2$ MIMO interference network. Since cooperation does not hurt, the outer bound for the $2\times 2\times 2$ MIMO interference network also applies to the $K\times K\times K$ SISO interference network.
The feasibility condition is restated as follows.

\begin{theorem}\label{sal}
(Rephrased from Theorem 2 of \cite{Ilan_Salman}) In order for a $K \times K \times K$ interference network to have $K$ DoF, we have the following two claims.
\begin{enumerate}
\item (Claim 1): If $\mathcal{S}_i$ is not connected to $\mathcal{R}_j, \forall i,j \in \{1,\cdots, K\}$, 
then the channel between all relays except $\mathcal{R}_j$ and all destinations except $\mathcal{D}_i$ must be rank-deficient.
\item  (Claim 2): As a dual statement, if $\mathcal{R}_i$ is not connected to $\mathcal{D}_j,  \forall i,j \in \{1,\cdots, K\}$, 
then the channel between all sources except $\mathcal{S}_j$ and all relays except $\mathcal{R}_i$ must be rank-deficient.
\end{enumerate}
\end{theorem}

This $K$-DoF feasibility condition perfectly fits the rank matching principle. In order to allow $K$ DoF in the network, if a certain link is not present (creating a rank-deficiency), then its paired  channel in the other hop must be rank-deficient as well. Let us show how  both claims follow from our Theorem \ref{out}. 

Without loss of generality consider Claim 1 when $i=j=1$. In order to map a $K \times K \times K$ interference network to a $2 \times 2 \times 2$ setting such that we can use Theorem \ref{out}, we allow full cooperation between all sources except $\mathcal{S}_1$ such that they become another super source that we call $\mathcal{S}_2'$. Similarly, all relays/destinations except $\mathcal{R}_1$/$\mathcal{D}_1$ are clustered to become a super relay/destination that we call $\mathcal{R}_2'$/$\mathcal{D}_2'$. With this transformation, Claim 1 becomes that in order for the $K \times K \times K$ interference network to have $K$ DoF, if $r(\mathcal{S}_1\mathcal{R}_1) = 0$, then $r(\mathcal{R}_2'\mathcal{D}_2') < K-1$. To prove this by contradiction, we show that if $r(\mathcal{S}_1\mathcal{R}_1) = 0$ and $r(\mathcal{R}_2'\mathcal{D}_2') = K-1$, the newly formed $2 \times 2 \times 2$ MIMO interference network can not have $K$ DoF, which in turn means that the original $K \times K \times K$ interference network can not have $K$ DoF as cooperation can never hurt the sum-DoF. 
So we wish to prove 
\begin{align}
r(\mathcal{S}_1\mathcal{R}_1) = 0, r(\mathcal{R}_2'\mathcal{D}_2') = K-1 \Rightarrow d_{\Sigma} < K.
\end{align}
For this purpose, let us substitute into (\ref{eq:main}) with $i=j=1, \bar{i} = \bar{j} = 2$, $r(\mathcal{S}_1\mathcal{R}_1) = 0, r(\mathcal{R}_2'\mathcal{D}_2') = K-1, M(\mathcal{S}_1) = M(\mathcal{D}_1) = M_r(\mathcal{R}_1) = M_t(\mathcal{R}_1) = 1, M(\mathcal{S}_2') = M(\mathcal{D}_2') = M_r(\mathcal{R}_2') = M_t(\mathcal{R}_2') = K-1$. Then we have
\begin{align}
d_{\Sigma} &\leq \frac{1}{2}[(1 + 1) + (K-1 + K -1)] - \left|0 - (K-1) - \frac{1}{2}[(1+1) - (K-1+K-1)]\right|\\
& = K - 1 < K,
\end{align}
Claim 2  which is the dual of Claim 1, similarly follows from Theorem \ref{out}, as Theorem \ref{out} itself has a dual character.

\section{Discussion}\label{sec:discussion}

Although the focus of this paper is primarily on the $2 \times 2 \times 2$ interference channel,  its fundamental nature  leads to broad applicability in general multiflow multihop networks, as evident from the various extensions considered in the previous section. Furthermore, note that the rank-matching bounds are not limited to wireless  networks. Indeed, as is the case with most DoF results, the same bounds are applicable to the deterministic counterparts of wireless networks over finite fields \cite{Krishnamurthy_Jafar_FiniteField, Hong_Caire, Jafar_TIM}. As such, they seem particularly useful to go beyond the Precoding-Based-Network-Alignment (PBNA) paradigm considered in \cite{Ramakrishnan_Das_Maleki_Markopoulou_Jafar_Vishwanath, Meng_Das_Ramakrishnan_Jafar_Markopoulou_Vishwanath}. In  PBNA a multiple unicast network is reduced to a single hop deterministic counterpart of a wireless interference network by allowing only linear operations (e.g., random linear network coding) at intermediate nodes, whereas all the intelligence lies at the source and destination nodes. As a step beyond PBNA one could allow some intelligence at a subset of the intermediate relay nodes.  For example, in a 2-unicast PBNA framework, (or a $K$-unicast setting which is reduced to 2-unicast by clustering of nodes) one could select 2 MIMO relay nodes, either because these nodes exist as such or by clustering, such that the network reduces to a $2\times 2\times 2$ layered MIMO interference network. Since the structure of the network is reflected in the rank deficiencies of the constituent channels, the rank-matching bounds are applicable and may lead to new insights.

\section{Proofs}
\subsection{Proof of Theorem \ref{out}}  \label{p1}

Consider the rank-mismatch bound (\ref{eq:main}) for $i=2, j=1$. It can be equivalently stated as the following two bounds.
\begin{eqnarray}
d_{\Sigma} &\leq& M({\mathcal{S}_2})  +M_r({\mathcal{R}_1}) +r({\mathcal{R}_2\mathcal{D}_1})-    r({\mathcal{S}_2\mathcal{R}_1})   \label{d11a}\\
d_{\Sigma} &\leq& M_t({\mathcal{R}_2}) + M({\mathcal{D}_1})  +   r({\mathcal{S}_2\mathcal{R}_1}) - r({\mathcal{R}_2\mathcal{D}_1}) \label{d11b} 
\end{eqnarray}

Consider (\ref{d11a}). Given a sequence of reliable coding schemes (indexed by $n$) spanning $n$ channel uses, we note that from ${\bf Y}^{2^n}_1, {\bf Y}^{2^n}_2, {\bf Y}^{3^n}_1$, one can decode both messages. From Fano's inequality, we proceed as follows.
\begin{eqnarray}
\allowdisplaybreaks
n(R_1 + R_2 - \epsilon)
&\leq& I(W_1,W_2; {\bf Y}^{2^n}_1, {\bf Y}^{2^n}_2, {\bf Y}^{3^n}_1) \\
&=& h({\bf Y}^{2^n}_1, {\bf Y}^{2^n}_2, {\bf Y}^{3^n}_1) - \underbrace{h({\bf Y}^{2^n}_1, {\bf Y}^{2^n}_2, {\bf Y}^{3^n}_1 | W_1,W_2)}_{\geq no(\log P)} \label{al0}\\
&\leq& h({\bf Y}^{2^n}_1) + h({\bf Y}^{3^n}_1|{\bf Y}^{2^n}_1) + h({\bf Y}^{2^n}_2 | {\bf Y}^{2^n}_1, {\bf Y}^{3^n}_1) + no(\log P) \\
&\leq& nM_r(\mathcal{R}_1) \log P + h({\bf Y}^{3^n}_1|{\bf Y}^{2^n}_1, {\bf X}^{2^n}_1) + h({\bf Y}^{2^n}_2 | {\bf Y}^{2^n}_1, {\bf Y}^{3^n}_1,W_1) \notag \\
&&~ + \underbrace{ I(W_1; {\bf Y}^{2^n}_2 | {\bf Y}^{2^n}_1, {\bf Y}^{3^n}_1)}_{=n o(n)} + no(\log P) \label{al1} \\
&\leq& nM_r(\mathcal{R}_1) \log P + h( {\bf H}^{2^n}_{12} {\bf X}^{2^n}_2 + {\bf Z}^{3^n}_1 | {\bf Y}^{2^n}_1, {\bf X}^{2^n}_1) \notag \\
&& ~+ h({\bf Y}^{2^n}_2 | {\bf Y}^{2^n}_1, {\bf Y}^{3^n}_1,W_1, {\bf X}_1^{1^n}) + no(\log P)  \label{al2} \\
&\leq& nM_r(\mathcal{R}_1) \log P + h( {\bf H}^{2^n}_{12} {\bf X}^{2^n}_2 + {\bf Z}^{3^n}_1) \notag \\
&& ~+ h({\bf H}^{1^n}_{22} {\bf X}^{1^n}_2 + {\bf Z}^{2^n}_2 |{\bf H}^{1^n}_{12} {\bf X}^{1^n}_2 + {\bf Z}^{2^n}_1, {\bf Y}^{3^n}_1, W_1, {\bf X}_1^{1^n}) + no(\log P)\label{al3} \\
&\leq& nM_r(\mathcal{R}_1) \log P + nr(\mathcal{R}_2\mathcal{D}_1) \log P \notag \\
&& ~+ h({\bf H}^{1^n}_{22} {\bf X}^{1^n}_2 + {\bf Z}^{2^n}_2 |{\bf H}^{1^n}_{12} {\bf X}^{1^n}_2 + {\bf Z}^{2^n}_1) + no(\log P) \label{al4} \\
&\leq& nM_r(\mathcal{R}_1) \log P + nr(\mathcal{R}_2\mathcal{D}_1) \log P \notag \\
&& ~+ n ~\text{rank} \left( \left[ 
\begin{array}{c}
{\bf H}^{1}_{22} \\
 {\bf H}^{1}_{12}
\end{array}
\right] \right)\log P - nr(\mathcal{S}_2 \mathcal{R}_1) \log P + no(\log P) \label{al5}\\
&\leq& n[ M_r(\mathcal{R}_1)+r(\mathcal{R}_2\mathcal{D}_1)+M(\mathcal{S}_2)-r(\mathcal{S}_2\mathcal{R}_1) ] \log P+ no(\log P) \label{alfin}
\end{eqnarray}
where the differential entropy of the second term in (\ref{al0}) is no less than the differential entropy of noise therein. In (\ref{al1}), the first term is a result of the fact that Gaussian distribution is the entropy maximizer subject to covariance constraint and ${\bf Y}_1^2$ has only $M_r(\mathcal{R}_1)$ dimensions, the second term follows from the fact that the transmitted signal of $\mathcal{R}_1$, ${\bf X}^{2^n}_1$ is a function of its received signal, ${\bf Y}^{2^n}_1$, and the fourth term is due to the property that from ${\bf Y}^{3^n}_1$, one can decode $W_1$. In (\ref{al2}), 
we subtract out the contribution of ${\bf X}_1^2$ from ${\bf Y}_1^3$ in the second term and use the property that ${\bf X}_1^1$ is a function of $W_1$ in the third term. In (\ref{al3}), the property that reducing conditioning can not increase entropy is used to get the second term and we subtract out the contribution of ${\bf X}_1^1$ from ${\bf Y}_2^2, {\bf Y}_1^2$ in the third term. In (\ref{al4}), the second term is due to the fact that rank$({\bf H}_{12}^2) = r(\mathcal{R}_2\mathcal{D}_1)$ and the third term is obtained by dropping conditioning, which can not increase entropy. (\ref{al5}) follows from the property that Gaussian distribution maximizes conditional entropy subject to covariance constraint and rank$({\bf H}_{12}^1) = r(\mathcal{S}_2\mathcal{R}_1)$. To obtain (\ref{alfin}), we use the fact that $$\text{rank} \left( \left[ 
\begin{array}{c}
{\bf H}^{1}_{22} \\
 {\bf H}^{1}_{12}
\end{array}
\right]\right)  \leq M(\mathcal{S}_2).$$

Finally, let first $n$ and then $P$ go to infinity. Then we normalize (\ref{alfin}) by $n \log P$ and arrive at (\ref{d11a}).

In fact, (\ref{d11b}) can also be shown similarly. However, let us provide an alternative proof that might be more intuitive.  To obtain this outer bound, we will give $\mathcal{D}_1$ certain side information through a genie such that $\mathcal{D}_1$ can decode both messages. 

First, we give $\mathcal{D}_1$ the part of the signal observed at $\mathcal{R}_1$ that is comprised only of the noise and what is sent from $\mathcal{S}_2$, that is, ${\bf S}_1 = {\bf H}_{12}^1 {\bf X}_2^1 + {\bf Z}_1^2$. Note that the ${\bf S}_1$ has no more than  $r(\mathcal{S}_2\mathcal{R}_1)$ DoF (prelog of differential entropy). Given any reliable coding scheme, $\mathcal{D}_1$ is assured to be able to decode $W_1$ and reconstruct the signal sent from $\mathcal{S}_1$, i.e., ${\bf X}_1^1$. Combined with ${\bf S}_1 $ and full channel knowledge, $\mathcal{D}_1$ is able to reconstruct the signal  observed by $\mathcal{R}_1$, that is ${\bf { Y}}_1^2 = {\bf H}_{11}^1{\bf X}_1^1 + {\bf S}_1 = {\bf H}_{11}^1{\bf X}_1^1 + {\bf H}_{12}^1{\bf X}_2^1 + {\bf Z}_1^2$. Then,  as we assume the encoding functions of the relays are globally known, $\mathcal{D}_1$ can construct the transmitted signal for $\mathcal{R}_1$, ${\bf X}_1^2$, by performing encoding on ${\bf { Y}}_1^2$ using the encoding function of $\mathcal{R}_1$.

Next, we give $\mathcal{D}_1$ the part of transmitted signal sent by $\mathcal{R}_2$ that \emph{is} seen at $\mathcal{D}_2$ but is \emph{not} seen at $\mathcal{D}_1$, that is ${\bf S}_2 = ({\bf H}_{22}^2/{\bf H}_{12}^2) {\bf X}_2^2 + {\bf Z}$, where ${\bf H}_{22}^2/{\bf H}_{12}^2$ consists of column vectors that span the intersection of the column-span of ${\bf H}_{22}^2$ and the null-space of ${\bf H}_{12}^2$. ${\bf Z}$ is independent noise distributed as ${\bf Z} \sim \mathcal{CN}(0,{\bf I})$.  
Because $\text{rank}({\bf H}_{12}^2) = r(\mathcal{R}_2\mathcal{D}_1)$ and $\text{rank}({\bf H}_{22}^2) \leq M_t({\mathcal{R}_2})$, 
the dimension of ${\bf S}_2$ is at most $M_t({\mathcal{R}_2}) - r(\mathcal{R}_2\mathcal{D}_1)$. As $\mathcal{D}_1$ knows ${\bf X}_1^2$, it can get the received signal sent from $\mathcal{R}_2$, ${\bf H}_{12}^2 {\bf X}_2^2 + {\bf Z}_1^3$ by subtracting the contribution of ${\bf X}_1^2$ from ${\bf Y}_1^3$. Thus, $\mathcal{D}_1$ now has access to $ {\bf X}_2^2$ within bounded noise-distortion (without loss of generality, eliminate redundant dimensions that are not seen by either destination, if any, from ${\bf X}_2^2$ and note that $({\bf H}_{22}^2/{\bf H}_{12}^2){\bf X}_2^2$ and ${\bf H}_{12}^2{\bf X}_2^2$ together provide sufficiently many linear equations to solve for  all the non-redundant dimensions of ${\bf X}_2^2$).

Now that it has access to ${\bf X}_1^2$, and ${\bf X}_2^2$ within bounded noise-distortion, $\mathcal{D}_1$ is able to construct (within bounded noise distortion) the total received signal observed at $\mathcal{D}_2$. 

As $\mathcal{D}_2$ is guaranteed to be able to decode $W_2$, so can $\mathcal{D}_1$ (possibly after reducing noise by a bounded amount that is inconsequential for DoF). Since, $\mathcal{D}_1$ is able to decode all messages from ${\bf Y}_1^3, {\bf S}_1, {\bf S}_2$, the sum-DoF of all messages is bounded by the sum of the dimensions (pre-logs) of ${\bf Y}_1^3, {\bf S}_1, {\bf S}_2$,
\begin{eqnarray}
d_\Sigma&\leq&M(\mathcal{D}_1)+r(\mathcal{S}_2\mathcal{R}_1)+M_t(\mathcal{R}_2)-r(\mathcal{R}_2\mathcal{D}_1)
\end{eqnarray}
which gives us (\ref{d11b}). 

Thus we have proved (\ref{eq:main}) for $i=2, j=1$. Since all bounds have the same structure, the proof applies for every choice of indices, $i,j\in\{1,2\}$.
\hfill$\Box$

\subsection{Proof of Theorem \ref{in}} \label{p2}
First, notice that the outer bound $\min(4r^{[1]}, 4r^{[2]}, 2M - |r^{[1]} - r^{[2]}|)$ is valid. The first two terms are min-cut max- flow bounds and the last term follows from Theorem \ref{out}.

As we will use linear schemes, which satisfy duality, we may assume $r^{[1]} \leq r^{[2]}$ without any loss of generality. In this case, the outer bound simplifies to $\min(4r^{[1]}, 2M - (r^{[2]} - r^{[1]}))$. 

\begin{figure}[h]
\center
\includegraphics[width=3.5in]{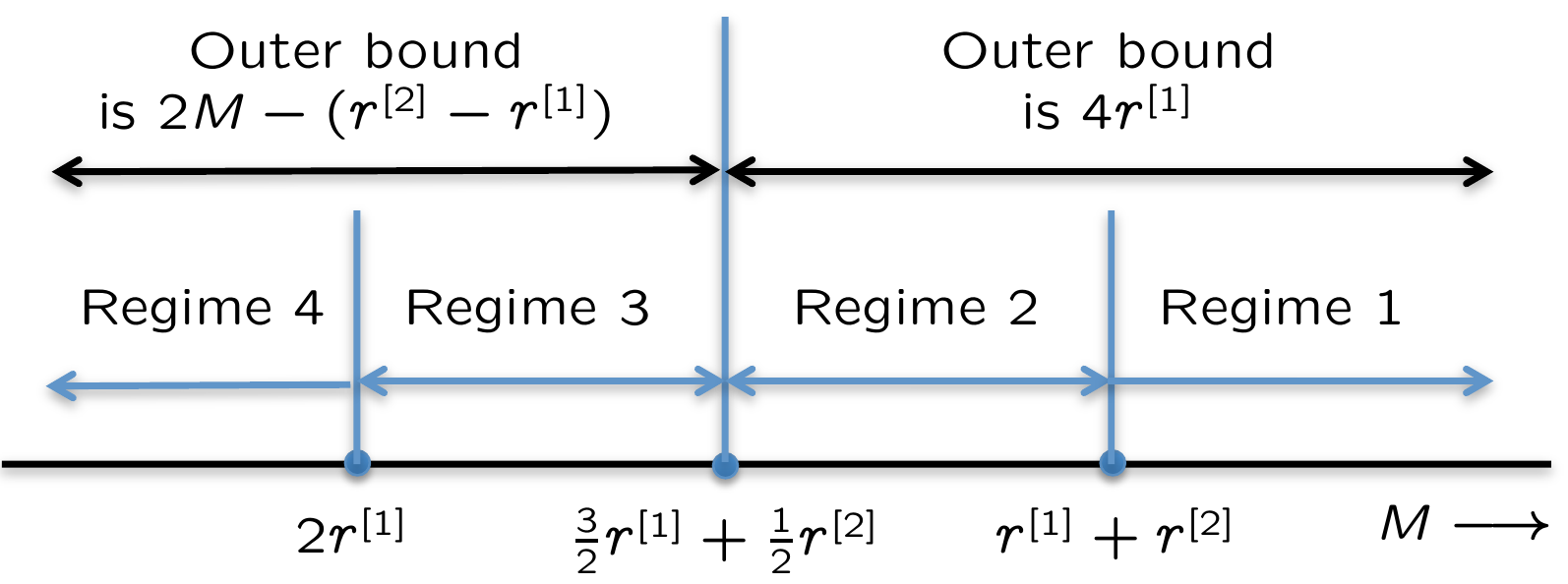}
\caption{\small The real axis is partitioned into 4 intervals, $(-\infty, 2r^{[1]}), (2r^{[1]}, \frac{3}{2}r^{[1]} + \frac{1}{2}r^{[2]}), (\frac{3}{2}r^{[1]} + \frac{1}{2}r^{[2]}, r^{[1]} + r^{[2]}), (r^{[1]} + r^{[2]}, +\infty)$. Depending on which interval  $M$ falls into, we have 4 regimes. For Regimes 1 and 2, the outer bound is $4r^{[1]}$ and for Regimes 3 and 4, the outer bound is $2M - (r^{[2]} - r^{[1]})$. Note that by the definition of rank, $M \geq r^{[2]} \geq r^{[1]}$, so we only consider those parameter regimes where this condition is true.}
\label{regime}
\end{figure}

For different configurations of $M, r^{[1]}, r^{[2]}$, both the outer bound and the channel constructed may vary. As such, based on  relationship between $M, r^{[1]}$ and $r^{[2]}$, we divide the total parameter space into 4 disjoint regimes (see Fig. \ref{regime}). We will first show for each regime,  that there exist channels that satisfy all rank-constraints, for which the outer bound is tight. We will conclude with the generalization that the bound is tight almost surely for generic channels.

\begin{itemize}
\item Regime 1 ($r^{[1]} + r^{[2]} \leq M$): The constructed channel appears in Fig. \ref{r1}. The connectivity is simple. The sources are connected to the relays with 4 orthogonal links. The relays are connected to the destinations with 4 orthogonal links and possibly a fully connected $2 \times 2$ subnetwork. For the channels that are shown as connected, one may choose the coefficients to be generic, that is, each non-zero channel coefficient is drawn independently from some continuous distribution bounded away from zero and infinity to avoid degenerate scenarios. For example, the first $r^{[1]}$ antennas of $\mathcal{S}_1$ are connected to the first $r^{[1]}$ antennas of $\mathcal{R}_1$ with a generic $r^{[1]} \times r^{[1]}$ (specifically, rank $r^{[1]}$) MIMO channel. We keep this assumption that every connected channel coefficient is generic for other regimes as well. Note that all rank conditions are satisfied.
Over such a channel, it is easy to achieve the outer bound, $4r^{[1]}$, as $\min(r^{[2]}, M - r^{[2]}) \geq r^{[1]}$ such that we can always route the messages over orthogonal links, by standard point to point MIMO capacity achieving schemes.

\begin{figure}[h]
\center
\includegraphics[width=4in]{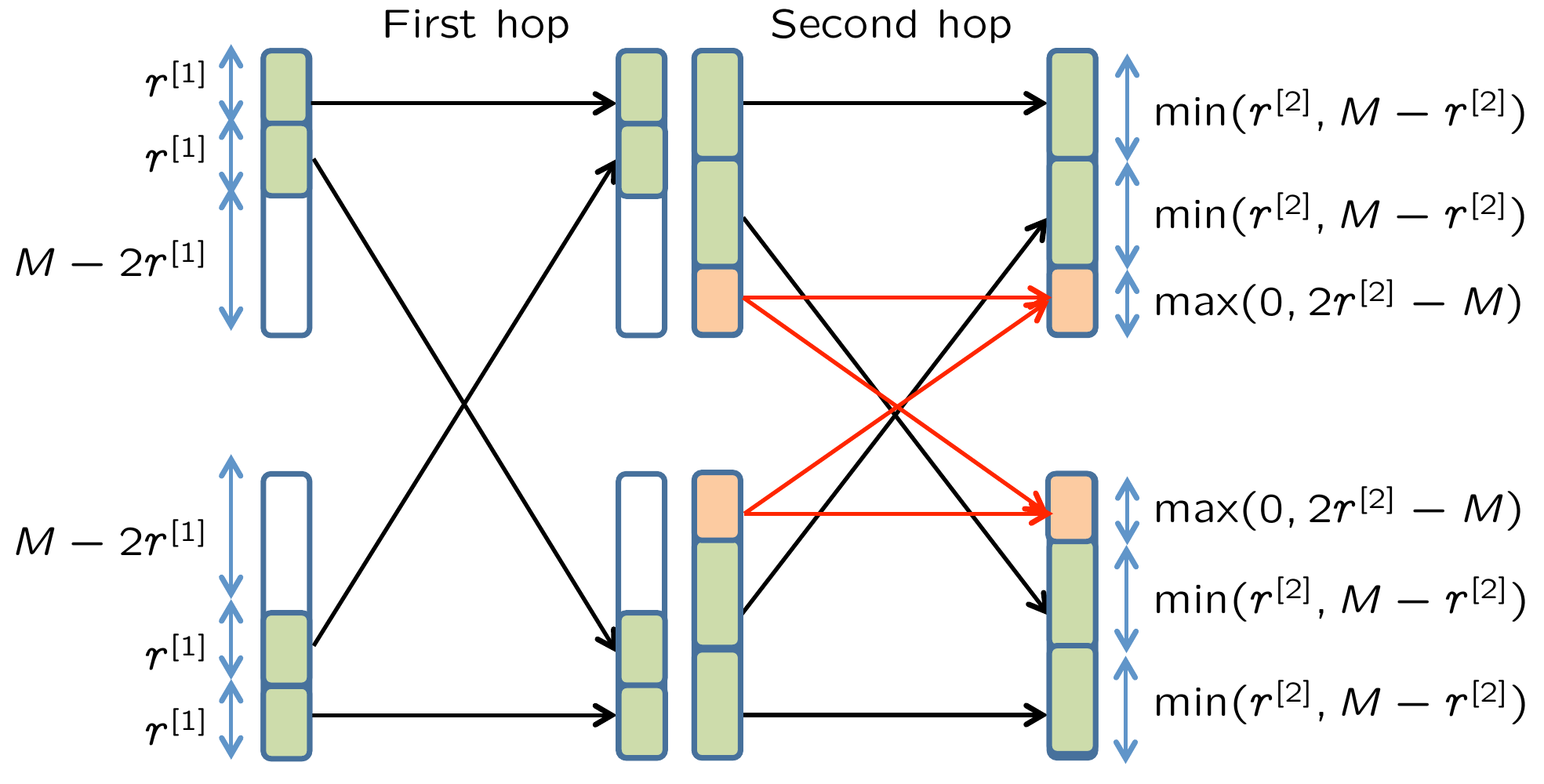}
\caption{\small Constructed channel for Regime 1. For clarity, the relay nodes are shown twice, one for the channels (receive side) of the first hop, the other for the channels (transmit side) of the second hop. 
}
\label{r1}
\end{figure}


\item{Regime 2} ($\frac{3}{2}r^{[1]} + \frac{1}{2}r^{[2]} \leq M < r^{[1]} + r^{[2]}$): The channel we construct is shown in Fig. \ref{r2}. The connectivity is same as Fig. \ref{r1}. The outer bound is still $4r^{[1]}$. In order to achieve that, pure routing will not suffice as each orthogonal link on the second hop only has DoF $M - r^{[2]}$, which can not support $r^{[1]}$ DoF, as in this regime, $r^{[1]} > M - r^{[2]}$. As a result, we have to use the fully connected $2 \times 2$ subnetwork on the second hop. The new idea here is viewing that as a $2 \times 2$ $X$ network with $2r^{[2]} - M$ antennas at each node, whose sum-DoF value is given by $\frac{4}{3}(2r^{[2]} - M)$ \cite{Jafar_Shamai}. Then as long as $4[r^{[1]} - (M - r^{[2]})]$, the total DoF that we fail to route to desired destinations, is smaller than $\frac{4}{3}(2r^{[2]} - M)$, we are able to utilize the interference alignment scheme over $X$ network to send the remaining $4[r^{[1]} - (M - r^{[2]})]$ DoF. We have
\begin{align}
4[r^{[1]} - (M - r^{[2]})] \leq \frac{4}{3}(2r^{[2]} - M) \Leftrightarrow 2M \geq 3r^{[1]} + r^{[2]}
\end{align}
which is satisfied in Regime 2. Therefore the scheme works.

\begin{figure}[h]
\center
\includegraphics[width=4in]{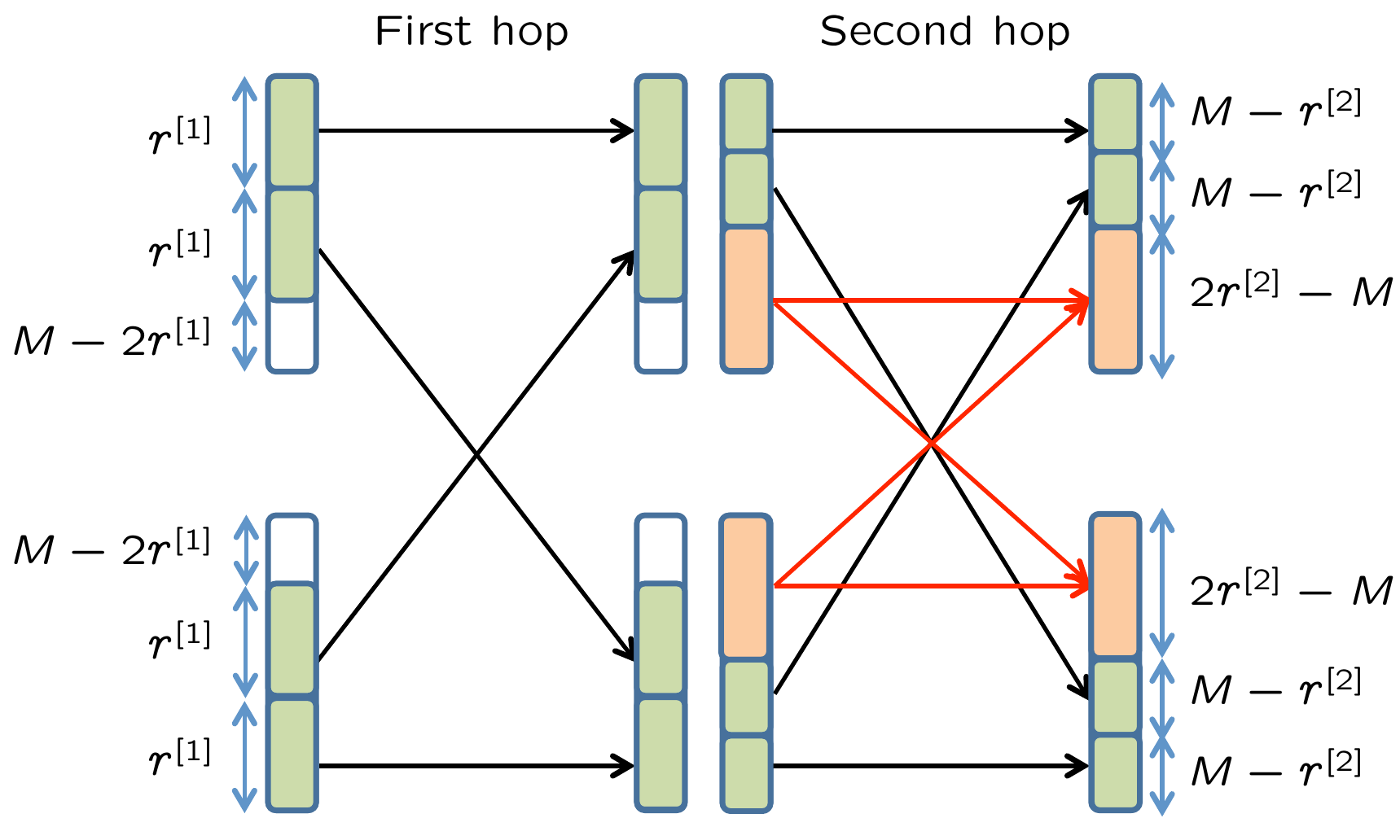}
\caption{\small Constructed channel for Regimes 2 and 3. The channel is almost the same that in Fig. \ref{r1}, where the only difference is that $M - r^{[2]}$ is smaller instead of bigger than $r^{[1]}$. To highlight such an important distinction which demands the use of $X$ scheme, we redraw the channel here.
}
\label{r2}
\end{figure}

\item Regime 3 ($2r^{[1]} \leq M <  \frac{3}{2}r^{[1]} + \frac{1}{2}r^{[2]}$): The channel is same as that used in Regime 2 (see Fig. \ref{r2}). Here the outer bound is $2M - (r^{[2]} - r^{[1]}) < 4r^{[1]}$. Note that in Regime 2, we have already saturated the fully connected $2 \times 2$ subnetwork by employing it as an $X$ network to the most. It may seem impossible to get something more. But thanks to the outer bound, we are not achieving $4r^{[1]}$ DoF, which means that the first hop has left capability. If we send same information from a source to both relays, the second hop can be employed as a broadcast channel (BC). Thus there exists a tradeoff, between employing the second hop as an $X$ network or a BC. $X$ scheme costs less on first hop but achieves fewer DoF on the second hop, while broadcast scheme achieves more DoF on the second hop but consumes more on the first hop. To determine the optimal ratio between them, we assume the second hop uses the $X$ scheme for $f_{X}$ fraction of time and the broadcast scheme for $f_{BC}$ fraction of time. Naturally, we have 
\begin{align}
f_X + f_{BC} = 1. \label{bcx1}
\end{align}

Note that for the fully connected $2 \times 2$ subnetwork, broadcast scheme has $2(2r^{[2]} - M)$ DoF and $X$ scheme has  $\frac{4}{3}(2r^{[2]} - M)$ DoF. Then by using $X$ scheme $f_X$ fraction of time and broadcast scheme $f_{BC}$ fraction of time, we need to have $2f_{BC}(2r^{[2]} - M) + \frac{4}{3}f_{X}(2r^{[2]} - M)$ DoF to send at the relays, which are received from the first hop. The broadcast messages need to be present at both relays and $X$ messages need only be at one relay, so we need to send a total of $4f_{BC}(2r^{[2]} - M) + \frac{4}{3}f_{X}(2r^{[2]} - M)$ DoF over the first hop, which should equal its capability, $4r^{[1]} - 4(M - r^{[2]})$. Note that $4(M-r^{[2]})$ DoF are occupied for routing messages to be sent over orthogonal links on the second hop.
Therefore, we have
\begin{align}
4f_{BC}(2r^{[2]} - M) + \frac{4}{3}f_{X}(2r^{[2]} - M) = 4r^{[1]} - 4(M - r^{[2]}). \label{bcx2}
\end{align}

Combining (\ref{bcx1})(\ref{bcx2}), we have
$$f_X = \frac{\frac{3}{2}(r^{[2]} - r^{[1]})}{2r^{[2]} - M}, f_{BC} = \frac{\frac{1}{2}(3r^{[1]}+r^{[2]} - 2M)}{2r^{[2]} - M},$$
such that the DoF value achieved by $X$ and broadcast schemes in total is $$2f_{BC}(2r^{[2]} - M) + \frac{4}{3}f_{X}(2r^{[2]} - M) = 3r^{[1]} + r^{[2]} - 2M + 2(r^{[2]}-r^{[1]}) =  r^{[1]} + 3r^{[2]} - 2M. $$
Adding up with $4(M-r^{[2]})$ routing DoF, we get $2M - (r^{[2]} - r^{[1]})$, as desired.

\item Regime 4 ($M  < 2r^{[1]}$): The constructed channel appears in Fig. \ref{r4}. We want to show that the outer bound, $2M - (r^{[2]} - r^{[1]})$, is achievable. The new element here is that the first hop itself contains a fully connected subnetwork. To utilize this, we pair it with the second hop to get a $2 \times 2 \times 2$ MIMO full rank interference channel with $2r^{[1]} - M$ antennas everywhere. By aligned interference neutralization (AIN), we achieve $2(2r^{[1]} - M)$ DoF \cite{Gou_Wang_Jafar_Jeon_Chung}. Then the fully connected subnetwork on the second hop is split into 2 parallel subnetworks. 
 Similar as before, we route $4(M - r^{[2]})$ DoF which saturates the orthogonal links on the second hop. We are left to use the fully connected $2 \times 2$ subnetwork with $2(r^{[2]} - r^{[1]})$ antennas at each node on the second hop. The first hop has unused DoF $4(M - r^{[1]}) - 4(M - r^{[2]}) = 4(r^{[2]} - r^{[1]})$, after AIN and routing. Here we also need to decide how to share the second hop with $X$ and broadcast schemes. Then following similar logic, we have
\begin{align}
f_{X} + f_{BC} &= 1 \\
\frac{8}{3} (r^{[2]} - r^{[1]})f_{X}  +  8(r^{[2]} - r^{[1]})f_{BC}  &= 4(r^{[2]} - r^{[1]})
\end{align}
from which we can solve $ f_X = \frac{3}{4}, f_{BC} = \frac{1}{4}$
such that the DoF value achieved is $r^{[2]} - r^{[1]}$ by broadcast scheme and $2(r^{[2]} - r^{[1]})$ by $X$ scheme. Adding up with those achieved by AIN and routing, we get $2(2r^{[1]} - M) + 4(M-r^{[2]}) + (r^{[2]} - r^{[1]}) + 2(r^{[2]} - r^{[1]}) = 2M - (r^{[2]} - r^{[1]})$, as desired.
\end{itemize}

\begin{figure}[h]
\center
\includegraphics[width=4in]{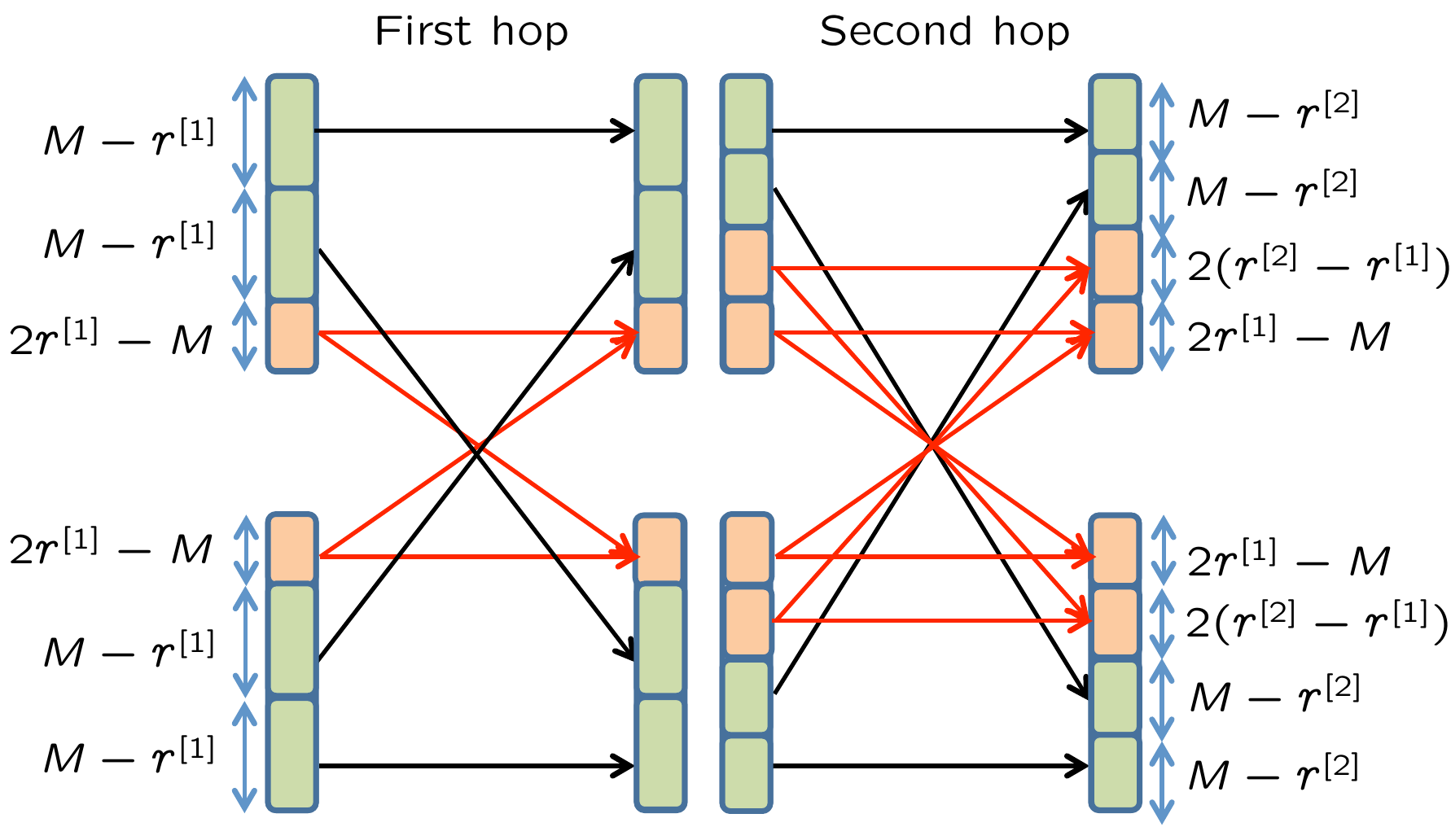}
\caption{\small Constructed channel for Regime 4.
}
\label{r4}
\end{figure}

As a summary, we list the achievable scheme used and corresponding DoF achieved in Table I. 

\begin{center}
\footnotesize
\begin{tabular}{|c|c|c|c|c|c|} \hline
\multicolumn{6}{ |c| }{Table I: DoF achieved by each scheme for each regime} \\ \hline
Regimes & AIN  & BC & $X$ & Routing & Total DoF \\ \hline
{$r^{[1]} + r^{[2]} \leq M$} & 0 & 0 & 0 & $4r^{[1]}$ & $4r^{[1]}$ \\ \hline 
{$\frac{3}{2}r^{[1]} + \frac{1}{2}r^{[2]} \leq M < r^{[1]} + r^{[2]}$} & 0  & 0  & {$4(r^{[1]} + r^{[2]} - M)$} & $4(M - r^{[2]})$ & $4r^{[1]}$ \\ \hline
{$2r^{[1]} \leq M <  \frac{3}{2}r^{[1]} + \frac{1}{2}r^{[2]}$} &  0 & $3r^{[1]}+r^{[2]}-2M$ & $2(r^{[2]}-r^{[1]})$ & $4(M-r^{[2]})$ & $2M-(r^{[2]}-r^{[1]})$ \\ \hline
{$M  < 2r^{[1]}$}  & $2(2r^{[1]}-M)$ & $r^{[2]}-r^{[1]}$  & $2(r^{[2]}-r^{[1]})$ & $4(M-r^{[2]})$ & $2M-(r^{[2]}-r^{[1]})$ \\ \hline
\end{tabular}
\normalsize
\end{center}

Finally, we consider fully generic channels, guided by insights from specific channel constructions presented for each of the regimes. In particular, we will show that through proper precoding, we can essentially create the specific channel constructed above such that the achievable scheme with DoF allocation as specified in Table I obtains the outer bound. Similarly, we have 4 regimes.

\begin{itemize}
\item Regime 1 ($r^{[1]} + r^{[2]} \leq M$): We consider the first hop. Referring to Fig. \ref{r1}, we want to create 4 orthogonal links, one from each source to each relay. Towards this end, we will choose 4 $M \times r^{[1]}$ precoding matrices, ${\bf V}_{ZF11}^1$, ${\bf V}_{ZF21}^1$, ${\bf V}_{ZF12}^1$ and ${\bf V}_{ZF22}^1$ as follows.
\begin{align}
{\bf V}_{ZF11}^1 \subseteq \mathcal{N} ({\bf H}_{21}^1), {\bf V}_{ZF21}^1 \subseteq \mathcal{N} ({\bf H}_{11}^1) \\
{\bf V}_{ZF12}^1 \subseteq \mathcal{N} ({\bf H}_{22}^1), {\bf V}_{ZF22}^1 \subseteq \mathcal{N} ({\bf H}_{12}^1)
\end{align}
where $\mathcal{N}({\bf A})$ denotes the right null space of matrix ${\bf A}$. Note that ${\bf V}^1_{ZFji}, i \in \{1,2\}, j \in\{1,2\}$ is used by $\mathcal{S}_i$, for $\mathcal{R}_j$ in the sense $\mathcal{R}_{\bar{j}}$ is zero forced. As the generic channel ${\bf H}_{ji}^1$ has rank $r^{[1]}$ such that $\dim(\mathcal{N} ({\bf H}_{ji}^1)) = M - r^{[1]}$ and $2r^{[1]} \leq r^{[1]} + r^{[2]} \leq M$, such ${\bf V}^1_{ZFji}$ exist. Moreover, at $\mathcal{S}_i$, the precoding matrix $[{\bf V}^1_{ZF1i} ~{\bf V}^1_{ZF2i}]$ has full rank as the two components are null spaces of generic channel matrices and the sum of their dimensions, $2r^{[1]}$ is smaller than the total space size, $M$. 
At $\mathcal{R}_j$, the receive signal space $[{\bf H}^1_{j1} {\bf V}^1_{ZFj1} ~ {\bf H}^1_{j2} {\bf V}^1_{ZFj2}]$ also has full rank as ${\bf H}^1_{ji} {\bf V}^1_{ZFji}$ is a subspace of ${\bf H}^1_{ji}$ and the column spaces of two generic matrices ${\bf H}^1_{j1}$ and ${\bf H}^1_{j2}$ (with rank $r^{[1]}$ each) do not intersect in an $M$ dimensional space, since $2r^{[1]} \leq M$. This process creates 4 orthogonal links.

The second hop is similar to the first hop. We choose precoding matrices at the relays such that undesired destination is zero forced. The linear independence of vectors of precoding matrix at the relay and receive signal space at the destination can be similarly proved. After creating such orthogonal links as in Fig. \ref{r1}, we can use routing to achieve the desired $4r^{[1]}$ DoF.

\item{Regime 2} ($\frac{3}{2}r^{[1]} + \frac{1}{2}r^{[2]} \leq M < r^{[1]} + r^{[2]}$): The first hop is same as Regime 1, using null spaces to create orthogonal links. On the second hop, $\mathcal{R}_i$ uses following precoding matrix ${\bf V}_i^2$ of size  $M \times 2r^{[1]}$.
\begin{eqnarray}
&& {\bf V}_{1}^2  = [{\bf V}_{ZF11}^2 \hspace{3mm} {\bf V}_{ZF21}^2 \hspace{3mm} {\bf V}_{X11}^2 \hspace{3mm} {\bf V}_{X21}^2] \\
&& {\bf V}_{2}^2  = [{\bf V}_{ZF12}^2 \hspace{3mm} {\bf V}_{ZF22}^2 \hspace{3mm} {\bf V}_{X12}^2 \hspace{3mm} {\bf V}_{X22}^2] \\
&& \dim({\bf V}_{ZFji}^2) = M-r^{[2]} \hspace{20mm} \\ 
&& \dim({\bf V}_{Xji}^2) = r^{[1]}+r^{[2]}-M \hspace{5mm} 
\end{eqnarray}
wherein ${\bf V}_{ZFji}^2 = \mathcal{N} ({\bf H}_{\bar{j} i}^2)$, and ${\bf V}_{Xji}^2$ are chosen such that the following $X$ network alignment conditions are satisfied.
\begin{eqnarray} \label{eqn:x_align}
{\bf H}_{11}^2 {\bf V}_{X21}^2 = - {\bf H}_{12}^2 {\bf V}_{X22}^2 \subseteq {\bf H}_{11}^2 \cap {\bf H}_{12}^2 \label{x1}\\
{\bf H}_{21}^2 {\bf V}_{X11}^2 = - {\bf H}_{22}^2 {\bf V}_{X12}^2 \subseteq {\bf H}_{21}^2 \cap {\bf H}_{22}^2 \label{x2}
\end{eqnarray}

Note that
\begin{equation}
\dim({\bf H}_{11}^2 \cap {\bf H}_{12}^2) = \dim({\bf H}_{21}^2 \cap {\bf H}_{22}^2) = 2r^{[2]} - M \geq {r^{[1]} + r^{[2]} - M} = \dim({\bf V}_{Xji}^2) 
\end{equation}
then ${\bf V}_{Xji}^2$ exist. With vectors chosen in this way, at $\mathcal{R}_i$, the precoding matrix ${\bf V}_i^2$ has $2r^{[1]} \leq M$ linear independent columns.
The signal space matrix at $\mathcal{D}_1$ is given as
\begin{eqnarray} 
[{\bf H}_{11}^2{\bf V}_1^2 \hspace{4mm} {\bf H}_{12}^2 {\bf V}_2^2] = [ {\bf H}_{11}^2 {\bf V}_{ZF11}^2 \hspace{4mm} {\bf H}_{12}^2 {\bf V}_{ZF12}^2 \hspace{4mm} {\bf H}_{11}^2 {\bf V}_{X11}^2 \hspace{4mm} {\bf H}_{12}^2 {\bf V}_{X12}^2 \hspace{4mm} {\bf H}_{11}^2 {\bf V}_{X21}^2]  
\end{eqnarray}
which has $2(M - r^{[2]}) + 3(r^{[1]} + r^{[2]} - M) = 3r^{[1]} + r^{[2]} - M \leq M$ vectors such that it has full rank, since the transmitted vectors are independent and  pass through  channels that are generic.
Similarly, the signal space matrix at $\mathcal{D}_2$ also has full rank. We can now use the first hop to transmit $4r^{[1]}$ DoF to the relays which then use a combination of zero forcing and $X$ scheme with precoding matrices as above to send these DoF to the destinations.

\item Regime 3 ($2r^{[1]} \leq M <  \frac{3}{2}r^{[1]} + \frac{1}{2}r^{[2]}$): The first hop is still the same and we have 4 orthogonal links with sum-DoF $4r^{[1]}$. According to Table I, to each relay, we will send $3r^{[1]} + r^{[2]} - 2M$ DoF of common message,  $(r^{[2]} - r^{[1]})$ DoF which will utilize $X$ scheme and $2(M-r^{[2]})$ DoF which will be sent by zero forcing, over the second hop. This is possible since $2(3r^{[1]} + r^{[2]} - 2M) + 2(r^{[2]} - r^{[1]}) + 4(M - r^{[2]}) = 4r^{[1]}$, which is supportable on the first hop. At each relay, the zero forcing and $X$ precoding vectors will be chosen the same as Regime 2. The precoding vectors for broadcast scheme are the same as $X$, by noting that for the solution of (\ref{x1})(\ref{x2}), if we are transmitting the same message out, the interference caused to the undesired destination is nulled (instead of aligned as in $X$ network). At each destination, the received signal consists of $2(M-r^{[2]})$ zero forcing vectors, $\frac{3}{2}(r^{[2]} - r^{[1]})$ $X$ beamformed vectors ($\frac{2}{3}$ of which are desired and the other $\frac{1}{3}$ interfering) and $\frac{1}{2}(3r^{[1]}+r^{[2]}-2M)$ broadcast vectors, for a total of $M$. Linear independency at the relays and destinations follow similarly.

\item Regime 4 ($M  < 2r^{[1]}$): On the first hop, in order to create the fully connected $2 \times 2$ sub-network as in Fig. \ref{r4}, we prove that there exist two $M \times (2r^{[1]} - M)$ matrices ${\bf U}_1^1, {\bf U}_2^1$ such that
\begin{align}
{\bf H}_{11}^1 {\bf U}_1^1 = {\bf H}_{12}^1 {\bf U}_2^1 \label{a1}\\
{\bf H}_{21}^1 {\bf U}_1^1 = {\bf H}_{22}^1 {\bf U}_2^1 \label{a2}
\end{align}
Note the difference with (\ref{x1}) (\ref{x2}) where the precoding vectors are different in the two equations.
For the solution of (\ref{a1}), the basis of ${\bf U}_1^1$ has rank $r^{[1]}$, $r^{[1]}-M$ of which will have ${\bf H}_{11}^1 {\bf U}_1^1 = 0$ and the remaining $2r^{[1]} - M$ will produce  
${\bf H}_{11}^1 {\bf U}_1^1 = {\bf H}_{11}^1 \cap {\bf H}_{12}^1$. Similarly, for the solution of (\ref{a2}), ${\bf U}_1^1$ has rank $r^{[1]}$. These two $r^{[1]}$ dimensional spaces will intersect in a $2r^{[1]} - M$ dimensional space, which is the solution that we seek since it satisfies both equations. Similar solution can be found for $ {\bf U}_2^1$ as well. Thus, we have found two $2r^{[1]} - M$ dimensional spaces, one at each relay, that are accessible by the same space at each source. This gives us a fully connected subnetwork.
Inside such a $2r^{[1]} - M$ dimensional space, we design an AIN solution as proposed in \cite{Gou_Wang_Jafar_Jeon_Chung}, where $\mathcal{S}_1$ sends $p \triangleq 2r^{[1]} - M$ symbols with $p$ precoding vectors ${\bf v}^1_{AIN1,1}, \cdots, {\bf v}^1_{AIN1,p}$ and $\mathcal{S}_2$ sends $2r^{[1]} - M - 1 = p-1$ symbols with $p-1$ precoding vectors ${\bf v}^1_{AIN2,1}, \cdots, {\bf v}^1_{AIN2, p-1}$. Each precoding vector has size $M \times 1$. The alignment relationship is same as that used in \cite{Gou_Wang_Jafar_Jeon_Chung} (see Table I of  \cite{Gou_Wang_Jafar_Jeon_Chung}). At $\mathcal{R}_1$, we have
\begin{align}
{\bf H}_{11}^1 {\bf v}^1_{AIN1,q+1} = {\bf H}_{12}^1 {\bf v}^1_{AIN2,q}, ~~~q = 1, \cdots, p-1 \label{ain1}
\end{align} 
and at $\mathcal{R}_2$
\begin{align}
{\bf H}_{21}^1 {\bf v}^1_{AIN1,q} = {\bf H}_{22}^1 {\bf v}^1_{AIN2,q}, ~~~q = 1, \cdots, p-1 \label{ain2}
\end{align}
Here to find a solution, we will start from a random 1 dimensional subspace of ${\bf U}_1^1$ and set it as ${\bf v}^1_{AIN1,1}$, then go through (\ref{ain1})(\ref{ain2}) to find all other vectors. Note that as $p = 2r^{[1]} - M$, we are guaranteed to find such independent vectors. By a similar aligned neutralization design on the second hop (see Table II of \cite{Gou_Wang_Jafar_Jeon_Chung}), we are able to send $2p-1 = 2(2r^{[1]}-M)-1$ DoF with AIN.
By considering a $k$-symbol extension, we can send $2k(2r^{[1]}-M)-1$ symbols over such symbol-extended network by AIN, resulting in $2(2r^{[1]} - M)$ DoF asymptotically.

All other symbols are sent by BC, $X$ and routing (over zero forced orthogonal links) as specified in Table I. The operations that create these equivalent channels are the same as Regime 3. This completes the description of the achievable scheme for generic channels.

\end{itemize}

\subsection{Proof of Theorem \ref{region}} \label{p-r}
As the DoF region in Theorem \ref{region} is symmetric in $r^{[1]}, r^{[2]}$ and we will use linear schemes, which satisfy duality, we may assume $r^{[1]} \leq r^{[2]}$ without loss of generality. In this case, the DoF region simplifies to
\begin{align}
d_1 + d_2 &\leq 2M - (r^{[2]} - r^{[1]}) \label{misterm}\\
d_1 &\leq \min(2r^{[1]},M) \label{cutterm1} \\
d_2 &\leq \min(2r^{[1]},M) \label{cutterm2}
\end{align}

\begin{figure}[h]
\begin{center}
\includegraphics[width=5.5in]{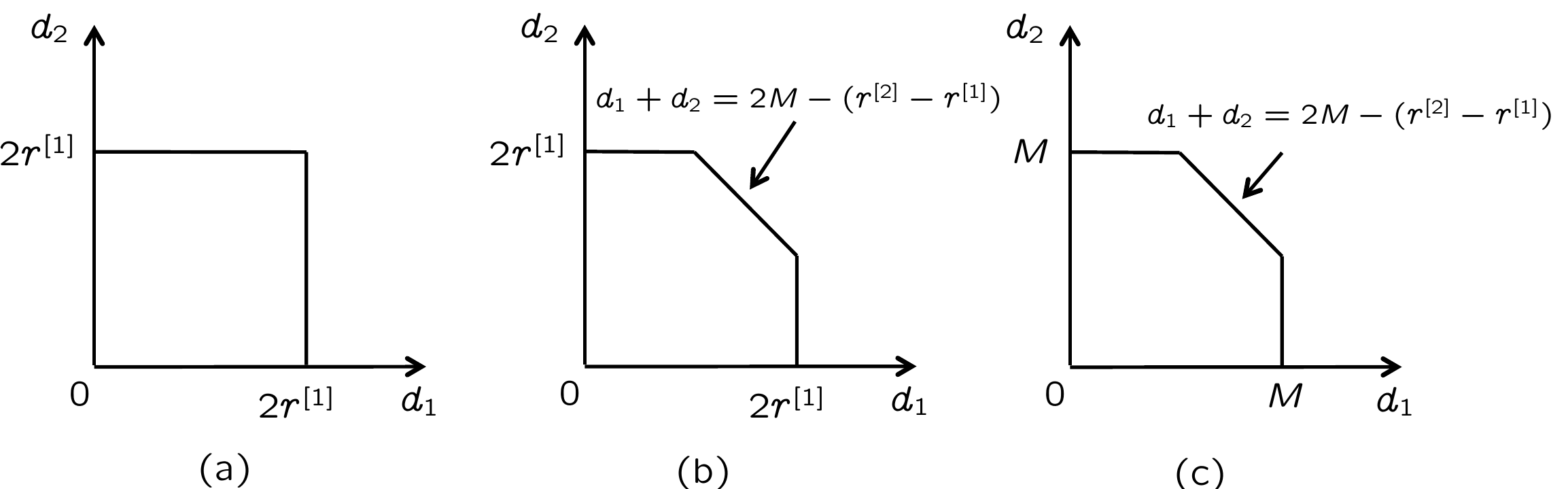}
\caption{\small The DoF region for the rank-constrained symmetric $2 \times 2 \times 2$ MIMO interference channel. (a) Regimes 1 and 2 ($\frac{3}{2}r^{[1]} + \frac{1}{2}r^{[2]} \leq M$), and (b) Regime 3  ($2r^{[1]} \leq M < \frac{3}{2}r^{[1]} + \frac{1}{2}r^{[2]}$), and (c) Regime 4  ($M < 2r^{[1]}$). }
\label{fig:region}
\end{center}
\end{figure}

Notice that (\ref{misterm}) is the rank-mismatch outer bound. (\ref{cutterm1}) and (\ref{cutterm2}) follow from the min-cut max-flow bounds.
Having proved the outer bound, we proceed to the achievability.
Similar to the sum-DoF case, for different parameter regimes, the DoF region varies. As such, we consider the same 4 regimes specified in Fig. \ref{regime}.
We have shown that with proper linear precoding, for each regime, we can create the specific constructed channel (see Fig. \ref{r1}, Fig. \ref{r2} and Fig. \ref{r4}) from generic channel matrices. Thus without loss of generality, we prove the DoF region of the constructed channel for each regime. The DoF region for each regime is plotted in Fig. \ref{fig:region}.

\begin{itemize}
\item Regimes 1 and 2 ($\frac{3}{2}r^{[1]} + \frac{1}{2}r^{[2]} \leq M$): The DoF region is a square as shown in Fig. \ref{fig:region}(a) and we only need to show the achievability of the corner point $(2r^{[1]}, 2r^{[1]})$, which is the sum-DoF optimal point that has been proved in Section \ref{p2}.

\item Regime 3  ($2r^{[1]} \leq M < \frac{3}{2}r^{[1]} + \frac{1}{2}r^{[2]}$): The DoF region is a pentagon as shown in Fig. \ref{fig:region}(b) and we want to show the achievability of the two corner points $(2r^{[1]}, 2M - r^{[1]} - r^{[2]}), (2M - r^{[1]} - r^{[2]}, 2r^{[1]})$. As the ranks of the channels 
are symmetric, it suffices to prove the achievability of DoF tuple $(d_1,d_2)  = (2r^{[1]}, 2M - r^{[1]} - r^{[2]})$. To this end, we use the same schemes introduced in Section \ref{p2}. The DoF allocation of each scheme is shown in Table II.

\begin{center}
\footnotesize
\begin{tabular}{|c|c|c|c|c|} \hline
\multicolumn{5}{ |c| }{Table II: $(d_1, d_2)$ achieved by each scheme for Regime 3} \\ \hline
$(d_1,d_2)$ & BC & $X$ & Routing & Total DoF \\ \hline
$d_1 $ & 0 & {$2(r^{[1]} + r^{[2]} - M)$} & $2(M - r^{[2]})$ & $2r^{[1]}$ \\ \hline 
$d_2$  & $3r^{[1]} + r^{[2]} -2M$ & $2(M - 2r^{[1]})$ & $2(M - r^{[2]})$  & $ 2M - r^{[1]} - r^{[2]}$\\ \hline
\end{tabular}
\normalsize
\end{center}

Recall that the constructed channel is shown in Fig. \ref{r2}. We wish to prove that over each hop, the channels can support the schemes in Table II. Over the first hop, $\mathcal{S}_1$ has two channels, one to $\mathcal{R}_1$ and one to $\mathcal{R}_2$. Each channel has DoF $r^{[1]}$ and can carry half of the DoF for $X$ scheme and half of the DoF for routing, as $r^{[1]} = (r^{[1]} + r^{[2]} - M) + (M - r^{[2]})$. Thus $d_1 = 2r^{[1]}$ DoF can be sent to the relays. Next we consider $d_2$. $\mathcal{S}_2$ has two channels to the relays, with total DoF $2r^{[1]}$ as well. Note that the messages to be sent with broadcast scheme need to be present at both relays and the messages to be sent with $X$ and routing schemes can be divided such that half of each appear in each relay. This is feasible since $r^{[1]} = (3r^{[1]} + r^{[2]} -2M) + (M - 2r^{[1]}) + (M - r^{[2]})$. This completes the proof of the first hop and we proceed to the second hop. Consider $d_1$, for the $2(r^{[1]} + r^{[2]} - M)$ DoF achieved by $X$ scheme, the interference alignment scheme in \cite{Jafar_Shamai} will guarantee that the interference caused at $\mathcal{D}_2$ has dimension $r^{[1]} + r^{[2]} - M$. This leaves enough space for the desired signal at $\mathcal{D}_2$, since $d_2 = 2M - r^{[1]} - r^{[2]} = M - (r^{[1]} + r^{[2]} - M)$. Therefore the desired signal can be decoded at $\mathcal{D}_2$. Similarly, the messages with $2(M- 2r^{[1]})$ DoF of $d_2$ that use $X$ scheme will occupy $M - 2r^{[1]}$ dimension at $\mathcal{D}_1$ and the messages that use broadcast and routing schemes will not be seen at $\mathcal{D}_1$. As such, $\mathcal{D}_1$ can decode the desired message as well, since $M = 2r^{[1]} + (M - 2r^{[1]}) = d_1 + (M - 2r^{[1]})$ such that the desired signal space and the interference space do not overlap.

\item Regime 4  ($M < 2r^{[1]}$): The DoF region is a pentagon as shown in Fig. \ref{fig:region}(c) and we only  need to show the achievability of the corner point $(M, M - (r^{[2]} - r^{[1]}))$, due to symmetry. To achieve that, the DoF allocation is shown in Table III.

\begin{center}
\footnotesize
\begin{tabular}{|c|c|c|c|c|c|} \hline
\multicolumn{6}{ |c| }{Table III: $(d_1, d_2)$ achieved by each scheme for Regime 4} \\ \hline
$(d_1,d_2)$ & AIN & BC & $X$ & Routing & Total DoF \\ \hline
$d_1 $ & $2r^{[1]} -M$ & 0 & {$2(r^{[2]} - r^{[1]})$} & $2(M - r^{[2]})$ & $M$ \\ \hline 
$d_2$  & $2r^{[1]} -M$ & $r^{[2]} - r^{[1]}$ & 0 & $2(M - r^{[2]})$ & $M-(r^{[2]} - r^{[1]})$\\ \hline
\end{tabular}
\normalsize
\end{center}

Recall that the constructed channel is shown in Fig. \ref{r4}. $2r^{[1]} -M$ DoF for each source will be sent by AIN over the fully connected $2 \times 2$ subnetwork of each hop. The first hop is able to send the remaining messages to the relays as it has left capability of $4(M - r^{[1]})$ DoF, which is equal to twice of the DoF of the messages to be sent with broadcast scheme, $r^{[2]} - r^{[1]}$, plus the DoF of the messages to be sent with $X$ and routing schemes, $2(r^{[2]} - r^{[1]}) + 4(M - r^{[2]})$. Next we consider the achievability of the messages sent with broadcast, $X$ and routing schemes over the second hop. The decoding at $\mathcal{D}_1$ is guaranteed since $\mathcal{D}_1$ does not see any interference. Interference caused by the messages sent with $X$ scheme of $d_1$ will occupy $r^{[2]} - r^{[1]}$ dimensions at $\mathcal{D}_2$, whose space do not overlap with its desired signal space (sent with broadcast scheme of DoF $r^{[2]} - r^{[1]}$) as the fully connected subnetwork has $2(r^{[2]} - r^{[1]})$ antennas. Therefore, $\mathcal{D}_2$ can decode the desired message as well. This completes the description of the achievable scheme for the DoF region.

\end{itemize}

\subsection{Proof of Theorem \ref{x}} \label{p-x}
The min-cut max-flow outer bound is trivial and we consider the achievability. As we will use linear schemes, which satisfy duality, we assume $r^{[1]} \leq r^{[2]}$ without loss of generality. Then the outer bound becomes $\min(4r^{[1]},2M)$. We still consider the 4 parameter regimes  in Fig. \ref{regime}. As linear precoding operation can reduce generic channel to the constructed channel, we need to prove the constructed channel only.

\begin{itemize}
\item Regimes 1 and 2 ($\frac{3}{2}r^{[1]} + \frac{1}{2}r^{[2]} \leq M$): In this case, the interference message setting can achieve the outer bound $4r^{[1]}$, so can the $X$ message setting as here we can use the interference channel scheme by setting the other two messages to be null. 

\begin{figure}[h]
\center
\includegraphics[width=3in]{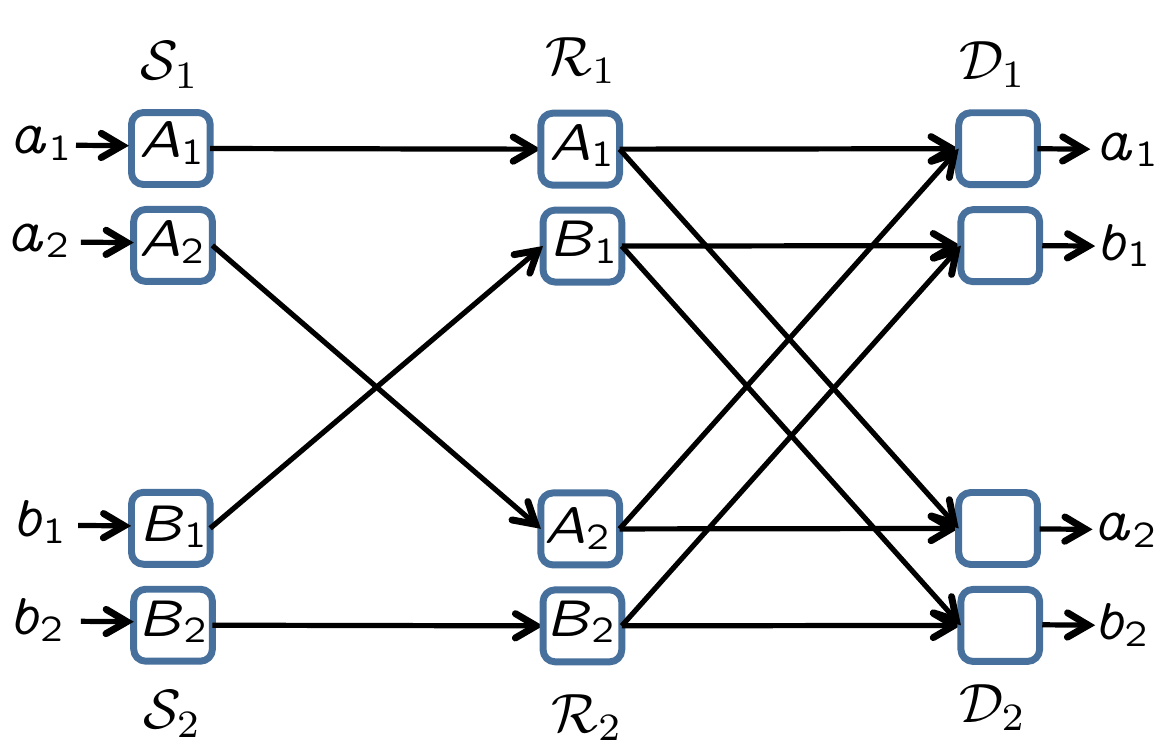}
\caption{\small Transmitted/Received symbols are shown inside the squares, which represent antennas. The relays use simple forwarding. $A_1, A_2$ denote two linear combinations of $a_1,a_2$ coded at $\mathcal{S}_1$ such that the interference caused by $a_1$ to $\mathcal{D}_2$ and the interference caused by $a_2$ to $\mathcal{D}_1$ are zero forced, over the second hop.
Similar coding is performed at $\mathcal{S}_2$ such that $b_1, b_2$ are received interference freely at $\mathcal{D}_1$ and $\mathcal{D}_2$, respectively.
}
\label{fig:x}
\end{figure}

\item Regime 3  ($2r^{[1]} \leq M < \frac{3}{2}r^{[1]} + \frac{1}{2}r^{[2]}$): Recall that the constructed channel appears in Fig. \ref{r2}. In order to achieve the outer bound $4r^{[1]}$, after routing $4(M - r^{[2]})$ DoF, we are left with $4(r^{[1]} + r^{[2]} - M)$ DoF, to be sent over the channel where the first hop consists of 4 orthogonal links with $r^{[1]} + r^{[2]} - M$ DoF each and the second hop is a fully connected $2 \times 2$ subnetwork with $2r^{[2]} - M$ antennas everywhere. As $2r^{[2]} - M \geq 2(r^{[1]} + r^{[2]} - M)$ in this regime, with zero forcing at the relays and the destinations, we are able to create $2(r^{[1]} + r^{[2]} - M)$ parallel fully connected $2 \times 2$ SISO subnetworks over the second hop. As such, we want to achieve $4(r^{[1]} + r^{[2]} - M)$ DoF over $r^{[1]} + r^{[2]} - M$ times the channel shown in Fig. \ref{fig:x}, which can be proved by showing that 4 DoF can be sent over the channel in Fig. \ref{fig:x}. We proceed to show this. In Fig. \ref{fig:x},  all nodes have 2 antennas, the first hop consists of 4 orthogonal links and the second hop consists of 2 parallel fully connected $2 \times 2$ subnetworks. In order to achieve 4 DoF, we wish to send 4 symbols over each channel use, where $a_1, a_2$ is sent from $\mathcal{S}_1$ to $\mathcal{D}_1$, $\mathcal{D}_2$, respectively and  $b_1, b_2$ is sent from $\mathcal{S}_2$ to $\mathcal{D}_1$, $\mathcal{D}_2$, respectively. With $a_1, a_2$ at $\mathcal{S}_1$, the transmitted symbols $A_1, A_2$ are designed such that with simple forwarding at the relays, the first antennas of $\mathcal{D}_1, \mathcal{D}_2$ will receive $a_1, a_2$ without interference, respectively. This is possible by precoding at $\mathcal{S}_1$, where global channel knowledge is known. With similar coding done at $\mathcal{S}_2$, $b_1$ can be sent to $\mathcal{D}_1$ and $b_2$ can be sent to $\mathcal{D}_2$, both interference freely. Therefore, 4 DoF is achievable here, as desired. 

\item Regime 4  ($M < 2r^{[1]}$): Recall that the constructed channel appears in Fig. \ref{r4}. The outer bound is $2M$. AIN achieves $2(2r^{[1]} - M)$ DoF and routing achieves $4(M - r^{[2]})$ DoF. The remaining $2M - 2(2r^{[1]} - M) - 4(M - r^{[2]}) = 4(r^{[2]} - r^{[1]})$ DoF can be sent over the remaining channel where the first hop consists of 4 orthogonal links with $r^{[2]} - r^{[1]}$ DoF each and the second hop can be reduced to $2(r^{[2]} - r^{[1]})$ parallel fully connected $2 \times 2$ SISO subnetwork, and this is $r^{[2]} - r^{[1]}$ times the channel shown in Fig. \ref{fig:x}. The achievability of 4 DoF over the channel in Fig. \ref{fig:x} is shown above and applying the scheme $r^{[2]} - r^{[1]}$ times achieves the desired remaining $4(r^{[2]} - r^{[1]})$ DoF. This completes the achievability proof.

\end{itemize}

\subsection{Proof of Theorem \ref{3hopin}} \label{p3}
We first consider some component channels and show in each case, the min-cut max-flow bounds are achievable. Then we consider the symmetric setting illustrated in Fig. \ref{3hop} and show that for arbitrary ranks of $r^{[1]}, r^{[2]}, r^{[3]}$, the channel can be decomposed into such component channels such that the min-cut max-flow bounds are achievable overall.

For all the component channels, we assume the connected channels are generic.
The first component channel is the 3 hop SISO fully connected interference channel, where the min-cut max-flow bound, 2 DoF are achievable \cite{Gou_Wang_Jafar_Jeon_Chung}, by cascading the first two hops to one single hop and employing the achievable scheme for the $2 \times 2 \times 2$ interference channel.

\begin{figure}[h]
\center
\includegraphics[width=4in]{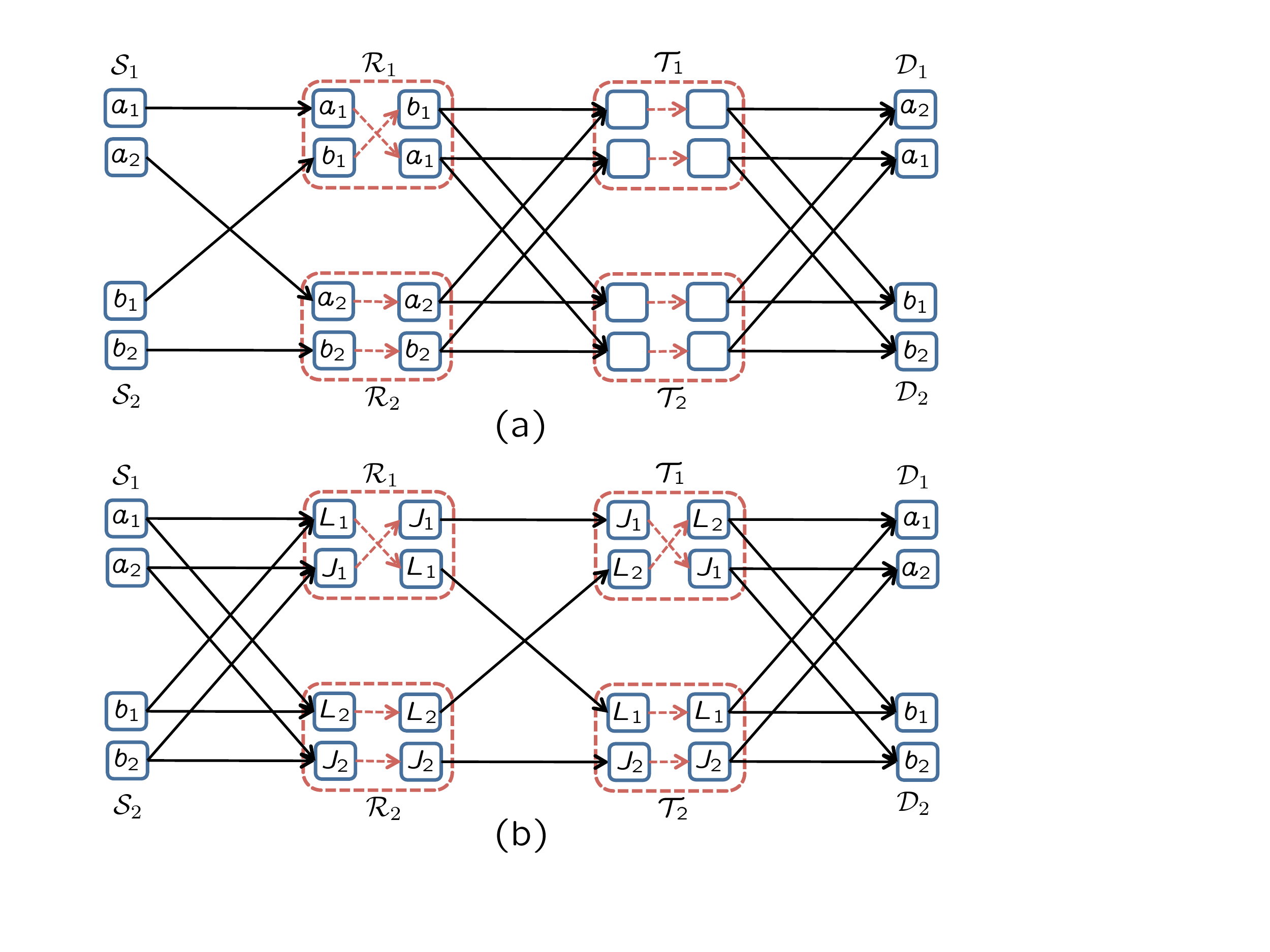}
\caption{\small The second class of component channels. Transmitted/Received symbols are shown inside the squares, which represent antennas. The relay nodes are shown twice (for both receiving and transmitting) and the linear coding inside the relay is shown by the dashed lines inside the dashed box. 
In this case, forwarding is sufficient to change the channel to parallel $2 \times 2 \times 2$ interference channels. In (b), $L_1, L_2$ each denotes a received linear combination of $a_1,b_1$ at the first antenna of $\mathcal{R}_1, \mathcal{R}_2$, respectively, and $J_1, J_2$ each denotes a received linear combination of $a_2, b_2$ at the second antenna of $\mathcal{R}_1, \mathcal{R}_2$, respectively.
}
\label{fig:2full}
\end{figure}

The second class of component channels is shown in Fig. \ref{fig:2full}, where all nodes have 2 antennas, one hop consists of 4 orthogonal links and the other hops consist of two parallel fully connected $2 \times 2$ subnetworks. We wish to show that the min-cut max-flow bound, 4 is achievable. For Fig. \ref{fig:2full}(a), after $\mathcal{S}_1, \mathcal{S}_2$ route 4 symbols to $\mathcal{R}_1, \mathcal{R}_2$, the last two hops become two parallel $2 \times 2 \times 2$ interference channels. The first $2 \times 2 \times 2$ interference channel consists of the first antennas of $\mathcal{R}_1, \mathcal{R}_2, \mathcal{T}_1, \mathcal{T}_2, \mathcal{D}_1, \mathcal{D}_2$. $\mathcal{R}_1$ wants to send $b_1$ to $\mathcal{D}_2$ and $\mathcal{R}_2$ wants to send $a_2$ to $\mathcal{D}_1$ (see Fig. \ref{fig:2full}(a)). Switching the destination indices will change this channel to the canonical $2 \times 2 \times 2$ interference channel such that 2 DoF can be achieved \cite{Gou_Wang_Jafar_Jeon_Chung}. The second antennas of $\mathcal{R}_1, \mathcal{R}_2, \mathcal{T}_1, \mathcal{T}_2, \mathcal{D}_1, \mathcal{D}_2$ form another $2 \times 2 \times 2$ interference channel where 2 DoF can be achieved \cite{Gou_Wang_Jafar_Jeon_Chung}. 
Next we consider Fig. \ref{fig:2full}(b).
The situation is similar. With the forwarding operation at the relays shown in Fig. \ref{fig:2full} (b), the first hop is connected to the third hop with two parallel $2 \times 2 \times 2$ interference channels and the min-cut max-flow bound, 4, is achievable.

\begin{figure}[h]
\center
\includegraphics[width=4in]{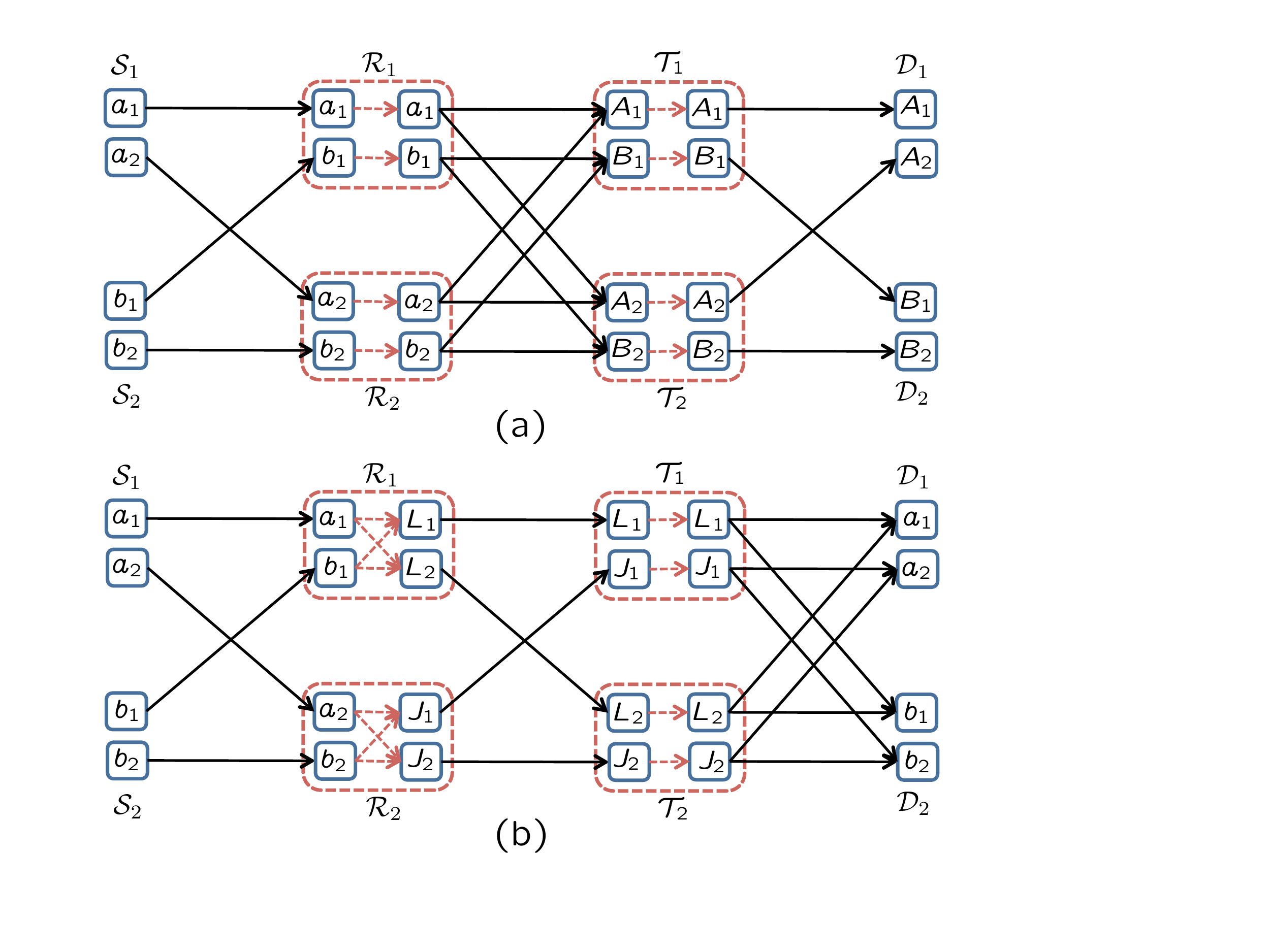}
\caption{\small The third class of component channels. In (a), $A_1,A_2$ are two received linear combinations of $a_1,a_2$ and $B_1,B_2$ are two received linear combinations of $b_1,b_2$, at corresponding antennas. In (b), $L_1, L_2$ denote two linear combinations of $a_1,b_1$ coded at $\mathcal{R}_1$ such that the interference caused by $b_1$ to $\mathcal{D}_1$ and the interference caused by $a_1$ to $\mathcal{D}_2$ are zero forced, over the last hop.
Similar coding is performed at $\mathcal{R}_2$ such that $a_2, b_2$ are received interference freely at $\mathcal{D}_1$ and $\mathcal{D}_2$, respectively.
}
\label{fig:1full}
\end{figure}

The third class of component channels is shown in Fig. \ref{fig:1full},  where all nodes have 2 antennas, two hops consist of 4 orthogonal links and the remaining hop consists of two parallel fully connected $2 \times 2$ subnetworks. The min-cut max-flow bound is 4 and we prove it is achievable.
The achievable scheme for Fig. \ref{fig:1full}(a) is easy. The generic channels do the necessary coding automatically and the relays just need to forward what they receive (see Fig. \ref{fig:1full}(a)). As such, $\mathcal{D}_1$ has two generic linear combinations of $a_1,a_2$ such that $\mathcal{D}_1$ is able to decode $a_1,a_2$ almost surely. Similarly, $\mathcal{D}_2$ can get 2 DoF almost surely, resulting in the achievability of 4 DoF.
The last two hops of Fig. \ref{fig:1full}(b) can be viewed as a dual of the last two hops of Fig. \ref{fig:1full}(a). With $a_1, b_1$ at $\mathcal{R}_1$, the transmitted symbols $L_1, L_2$ are designed such that over the last hop, $\mathcal{D}_1$ receives $a_1$ and $\mathcal{D}_2$ receives $b_1$, both interference freely. This coding is possible because global channel knowledge is available at the relays, specifically $\mathcal{R}_1$ knows the channels of the last hop. Note that this mixing operation at $\mathcal{R}_1$ is necessary and non-trivial. It is guided by the rank-matching principle such that the first two hops would appear as fully connected, to match the third hop.
Similar operation is done at $\mathcal{R}_2$. Therefore 4 DoF are achievable almost surely.

\begin{figure}[h]
\center
\includegraphics[width=4in]{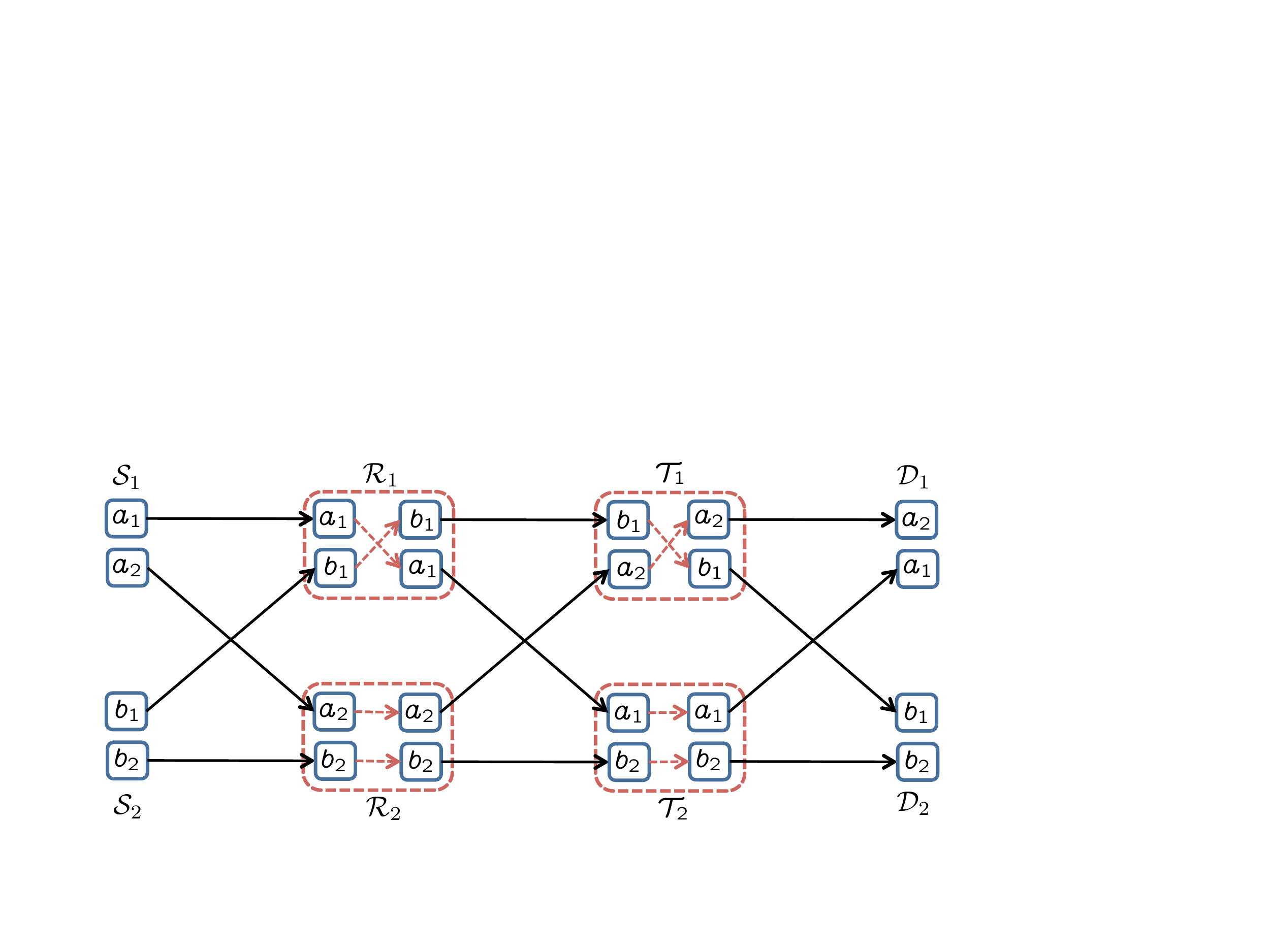}
\caption{\small The fourth component channel. Routing over 4 disjoint paths achieves 4 DoF.}
\label{fig:allorth}
\end{figure}

The fourth component channel is shown in Fig. \ref{fig:allorth}, where all nodes have 2 antennas and each hop consists of 4 orthogonal links. A routing solution achieves 4 DoF, the min-cut max-flow bound. 

Next we proceed to consider the symmetric setting. As we will use linear schemes, which satisfy duality, we assume $r^{[1]} \leq r^{[3]}$ without loss of generality. 

\begin{figure}[h]
\center
\includegraphics[width=2.5in]{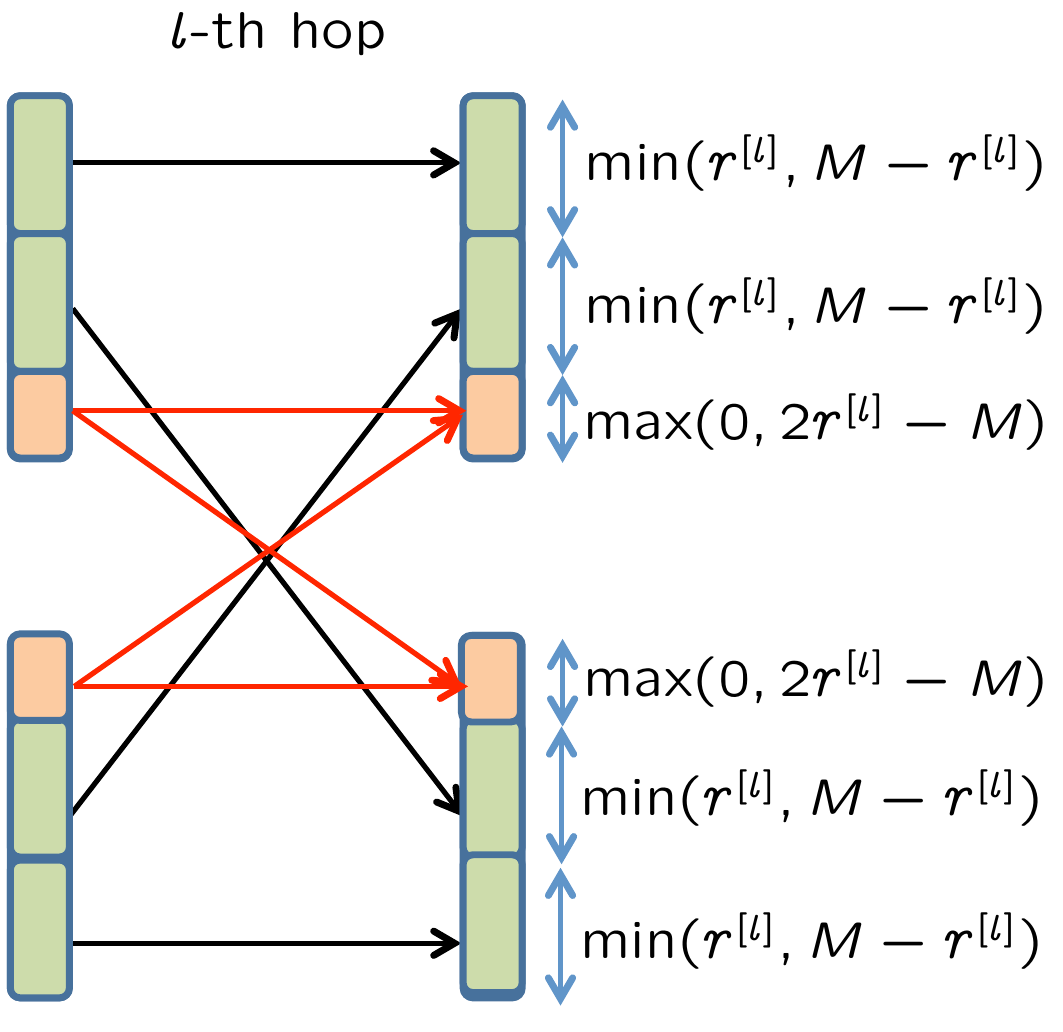}
\caption{\small The virtual channel created by linear precoding.}
\label{virtual}
\end{figure}

Using the same linear precoding techniques as in Section \ref{p2}, for any hop, we are able to create a virtual channel as shown in Fig. \ref{virtual}, which consists of 4 orthogonal links and possibly a fully connected $2 \times 2$ subnetwork, with corresponding dimensions. As such, we will first exploit the first component channel by cascading fully connected subnetworks over 3 hops. After exhausting this capability, one hop is left with no fully connected subnetwork, then we use the second class of component channels where two hops still have fully connected $2 \times 2$ subnetworks. Note that as $r^{[1]} \leq r^{[3]}$, we have only two cases, corresponding to the two shown in Fig. \ref{fig:2full}. After exhausting the second class, we use the third class of component channels where only 1 hop has some left fully connected $2 \times 2$ subnetwork, as shown in Fig. \ref{fig:1full}. Finally, we turn to the fourth component channel where all links are orthogonal. Note that the 4 classes of component channels are spatial scale invariant, meaning that if we scale the number of antennas and the ranks of each channel by a common factor, the total DoF will scale by the same factor.
Therefore each component channel achieves the min-cut max-flow bound and the comprised channel will also achieve the min-cut max-flow bound. This completes the proof.

\bibliographystyle{IEEEtran}
\bibliography{Thesis}
\end{document}